\documentclass[superscriptaddress,nofootinbib,preprintnumbers,rmp,letterpaper]{revtex4}
\usepackage{amsmath}
\usepackage{amsfonts}
\usepackage{amssymb}
\usepackage{epsfig}

\begin{document}

\title{Demagnetization via Nucleation of the Nonequilibrium Metastable Phase
in a Model of Disorder}
\author{Pablo I. Hurtado}
\email{phurtado@onsager.ugr.es}
\affiliation{Laboratoire des Collo\"{\i}des, Verres et Nanomat\'{e}riaux, Universit\'{e}
Montpellier II, Montpellier 34095, France}
\affiliation{Instituto \textquotedblleft Carlos I\textquotedblright\ de F\'{\i}sica Te\'{o}rica y 
Computacional, and Departamento de Electromagnetismo y F\'{\i}sica de la Materia, 
Universidad de Granada, E-18071 Granada, Espa\~{n}a}
\author{J. Marro$^{2}$}
\author{P.L. Garrido$^{2}$}

\begin{abstract}
We study both analytically and numerically demagnetization via nucleation of
the metastable phase in a two--dimensional nonequilibrium Ising ferromagnet
at temperature $T$. Canonical equilibrium is dynamically impeded by a weak
random perturbation which models homogeneous disorder of undetermined
source. We present a simple theoretical description, in perfect agreement
with Monte Carlo simulations, assuming that the decay of the nonequilibrium
metastable state is due, as in equilibrium, to the competition between the
surface and the bulk. This suggests one to accept a nonequilibrium \textit{%
free--energy} at a mesoscopic/cluster level, and it ensues a nonequilibrium 
\textit{surface tension} with some peculiar low--$T$ behavior. We illustrate
the occurrence of intriguing nonequilibrium phenomena, including: (\textit{i}) 
stochastic resonance phenomena at low $T$ which stabilize the metastable state as 
temperature increases; (\textit{ii}) reentrance of the limit of metastability under strong
nonequilibrium conditions; and (\textit{iii}) noise--enhanced propagation of
domain walls. The cooperative behavior of our system, which is associated to
the interplay between thermal and nonequilibrium fluctuations, may also be
understood in terms of a Langevin equation with additive and multiplicative
noises. We also studied metastability in the case of open boundaries as it
may correspond to a magnetic nanoparticle. We then observe the most
irregular relaxation triggered by the additional surface randomness. In
particular, at low $T,$ the relaxation becomes discontinuous as occurring by
way of scale--free avalanches, so that it resembles the type of relaxation
reported for many complex systems. We show that this results from the
superposition of many demagnetization events, each with a well--defined
scale which is determined by the curvature of the domain wall at which it
originates. This is an example of (apparent) scale invariance in a
nonequilibrium setting which is not to be associated with any familiar kind
of criticality.\smallskip

\end{abstract}

\maketitle

\section{\label{intro}Introduction}

Many different natural phenomena involve metastable states that, eventually,
decay via nucleation. Some familiar examples appear in the flow of electrical current through resistors, 
relaxation in amorphous materials, glasses and gels, domain wall motion in hysteretic disordered magnets, 
granular media evolution, earthquake dynamics, protein conformations, or false vacuum states in quantum 
field theory. There is a great amount of information concerning these situations but a
full microscopic theory of metastability and nucleation is elusive 
\cite{Langer2,Penrose2,Penrose1,Gunton1,Langer1,Rikvold,Debenedetti,FDT,Franz}. 
To begin with, there are two main coupled difficulties. One is that
metastability concerns dynamics \cite{Penrose1,Gunton1}. The systems of
interest typically show a complex free--energy landscape with (many) local minima, which
are metastable in the sense that they trap the system for a long time.
One may imagine that, eventually, relaxation occurs when the system after
long wandering finds a proper path between the minima. This results in
a complicate coupling of dynamics and thermodynamics \cite{Stillinger}. A second
difficulty is that many systems of interest cannot reach thermal equilibrium even after relaxation.
In general, they are open to the environment, which often induces currents
of matter or energy, or they are subject to agents which impose opposing
tendencies which typically break detailed balance \cite{MarroDickman}. These perturbations result in a final 
steady state which cannot be described by a Gibbsian measure, i.e. a nonequilibrium stationary state. Consequently, 
thermodynamics and ensemble statistical mechanics do not hold in these systems, which is a serious drawback.

These difficulties make the field most suitable for simple--model analysis.
Indeed, the (two--dimensional) kinetic Ising model has been the subject of
many studies of metastability not only in the case of periodic boundaries \cite{ns,sch,transfer,Rikvold}
but also for finite lattices with free boundaries \cite{nico,Vacas,contour,Cirillo}. 
The latter try to capture some of the physics of demagnetization in very
dense media where magnetic particle sizes typically range from mesoscopic down to
atomic levels. If one keeps oneself away from specific models, metastable
states are often treated in the literature as \textit{rare} equilibrium
states, at least for times much shorter that the relaxation time, and it has
been shown that one may define a metastable state in a properly constrained
(equilibrium) ensemble \cite{Penrose2,Penrose1,Franz}, and that most
equilibrium concepts may easily be adapted \cite{Gunton1,Langer1,Rikvold,Debenedetti,FDT}.

In this paper we present a detailed study of metastability (and nucleation)
in a \textit{nonequilibrium} model. In order to deal with a simple
microscopic model of metastability, we study a two--dimensional kinetic
Ising system, as in previous studies. However, for the system to exhibit
nonequilibrium behavior, time evolution is defined here as a superposition
of the familiar thermal process at temperature $T$ and a weak
completely--random process. This competition is probably one of the simplest, both
conceptually and operationally, ways of impeding equilibrium. Furthermore,
one may argue that it captures some underlying disorder induced by random
impurities or other causes which are unavoidable in actual samples. The
specific origin for such dynamic randomness will vary with the situation
considered. We mention that a similar mechanism has already been used to
model the macroscopic consequences of rapidly--diffusing local defects \cite%
{MarroDickman} and quantum tunneling \cite{Vacas} in magnetic materials, for
instance.

The weakest perturbation of that kind happens to modify essentially the
canonical equilibrium. We observe, however, some \textit{structural
similarity} at the mesoscopic level which suggests that cluster dynamics is,
as in equilibrium, the consequence of a competition between the surface and
the bulk. Therefore, we postulate the existence of a nonequilibrium \textit{%
free--energy} and an associated nonequilibrium \textit{surface tension.}
These functions are formally similar to the ones in equilibrium but behave
peculiarly due to the different nature of the system states. In fact, two
main predictions follow that are counter intuitive. One is that, at low $T$, the
stability of the metastable state is enhanced by increasing temperature, 
as a consequence of resonance between thermal and nonequilibrium noises. The second main
prediction is that the limit of metastability or \textit{pseudospinodal}, which
separates the metastable phase from the unstable one, exhibits reentrant
behavior as a function of $T$ for strong enough nonequilibrium conditions.
Our predictions are nicely confirmed in related Monte Carlo (MC)
simulations. We also show that the most important aspects of the model
behavior are contained in a Langevin--like description which contains a
multiplicative noise in addition to the more familiar additive one.

The theory above concerns an infinite system, and the related computer
simulations are, therefore, for lattices with periodic boundary conditions.
However, we also studied other types of boundary conditions, including free
borders. Previous work showed that free borders, which are most relevant
when dealing with magnetic nanoparticles, for instance, determine
importantly the \textit{particle} demagnetization \cite{contour,Cirillo}. We
here confirm this and observe that, for free borders, but not for periodic
boundary conditions, demagnetization occurs, via \textit{avalanches, }%
through many different metastable--like configurations. These avalanches
exhibit power--law size and lifetime distributions in a way that closely
resembles the relaxation in many cases in nature. We show that this is a
consequence in the model of the superposition of different events, each with
a typical scale which is determined by the curvature of the interface at
which the avalanche originates.

The paper is organized as follows. We define the model in the next section.
In section \ref{sect3}, we set the basis of our approach. In particular, $\S 
$\ref{31} concerns a mean--field approximation which allows for a discussion
in $\S $\ref{32} of the nonequilibrium surface tension. The next subsection
introduces our ansatz for a nonequilibrium \emph{free--energy} cost of 
excitations in the metastable phase, and we test the ensuing predictions against MC
simulations. In $\S $\ref{34} we present a Langevin theory which is also a
good description of the nonequilibrium situation of interest here. $\S $\ref%
{35} is devoted to study the limit of metastability. In section \ref{sect4}
we study the case of open boundaries. The statistics of the resulting
avalanches are analyzed in $\S $\ref{41}, and $\S $\ref{42} and $\S $\ref{43}
are devoted to discussions, respectively, of our observations and of its
possible extension to interpret avalanche data from actual systems. Finally, 
section $\S $\ref{conc} contains a brief conclusion. Some technical details are left for
the appendices.

\section{The Model\label{model}}

Consider a two--dimensional square lattice of side $L$ and, for the moment,
periodic boundary conditions (or, eventually, the thermodynamic limit
condition $L\rightarrow \infty ).$ We define a \textit{spin} variable $%
s_{i}=\pm 1$ at each node, $i\in \lbrack 1,N\equiv L^{2}].$ Spins interact
among them, and with an external magnetic field $h,$ via the Ising
Hamiltonian function 
\begin{equation}
\mathcal{H}\left( \mathbf{s}\right) =-\sum_{\left\vert i-j\right\vert
=1}s_{i}s_{j}-h\sum_{i=1}^{N}s_{i}\,,  \label{hamilt}
\end{equation}%
where $\mathbf{s=}\left\{ s_{i}\right\} $ and the first sum runs over all
nearest--neighbors (NN) pairs. Time evolution proceeds by stochastic
dynamics consisting of single--spin flips with the transition rate 
\begin{equation}
\omega (\mathbf{s}\rightarrow \mathbf{s}^{i})=p+(1-p)\Psi (\beta \Delta 
\mathcal{H}_{i})\,,  \label{rate}
\end{equation}%
where $0<p<1$ is a parameter, $\mathbf{s}$ and $\mathbf{s}^{i}$ stand,
respectively, for the system configuration before and after flipping the
spin at node $i$, $\Delta \mathcal{H}_{i}$ is the \textit{energy} cost of
the flip, and $\beta =1/T.$ The (arbitrary) function in (\ref{rate}) is set $%
\Psi (x)=\text{e}^{-x}(1+\text{e}^{-x})^{-1},$ corresponding to the
so--called \textit{Glauber rate}, in our simulations. However, we shall also
refer below to the case $\Psi (x)=\text{min}[1,\text{e}^{-x}]$ which
corresponds to the \textit{Metropolis rate.}

For any $p,$ two different heat baths compete in (\ref{rate}): One is at
temperature $T$ and operates with probability $(1-p)$, while the other
induces completely--random (\emph{infinite} temperature) spin flips with
probability $p$. As a consequence of this competition, a nonequilibrium
steady state sets in asymptotically which cannot be characterized by a
Gibbsian invariant measure \cite{Pedro1,MarroDickman}. In particular, the
rate (\ref{rate}) violates detailed balance for any $p>0.$ However, the
system reduces for $p=0$ to the familiar Ising model with rate $\Psi $ which
goes asymptotically to the equilibrium state for temperature $T$ and Hamiltonian 
$\mathcal{H}$. This model is a particular case of a more general class of systems, 
characterized by a coupling to a number of different heat baths leading to nonequilibrium 
behavior, whose static properties have received considerable attention in recent 
years \cite{MarroDickman}.

\begin{figure}[t]
\centerline{
\psfig{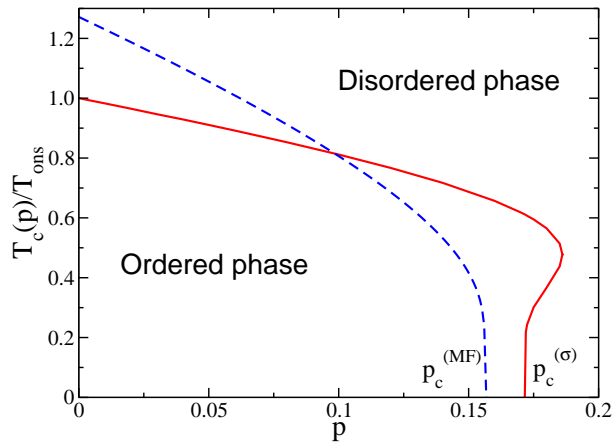}}
\caption{{\protect\small (Color) Phase diagram for the nonequilibrium Ising
ferromagnet of interest here. The solid (red) line corresponds to the
critical temperature as obtained from the nonequilibrium \emph{surface
tension} in $\S $\protect\ref{32}. The order disappears at $T=0$ for $%
p>p_{c}=(\protect\sqrt{2}-1)^{2}\approx 0.17$. Notice, however, the
reentrant \emph{blob} for $p_{c}<p<p_{c}^{\ast }\approx 0.186$ and
intermediate temperature. The dashed (blue) line is derived from a
first-order (pair approximation) mean--field theory, namely, eq. (\protect
\ref{TcMF}) in $\S $\protect\ref{31}. }}
\label{pd}
\end{figure}

For zero magnetic field, the model exhibits an order--disorder continuous phase transition 
at critical temperature $T_{c}(p)<T_{c}(p=0)\equiv T_{\text{ons}}$, where 
$T_{\text{ons}}=2/\ln (1+\sqrt{2})$ is the Onsager temperature. Fig. \ref{pd} 
shows the model phase diagram $T_{c}(p)$ as obtained by two
different approaches, as described in $\S $\ref{31} and $\S $\ref{32},
respectively. In the former case, corresponding to a first-order (pair) mean--field
approximation, the predicted equilibrium ($p=0$) critical temperature is the
Bethe temperature $T_{\text{bethe}}=4/\ln 4\approx 1.27T_{\text{ons}}$, and
order disappears at $T=0$ for $p>p_{c}^{(MF)}=5/32\approx 0.156$. A more
accurate estimation follows from the nonequilibrium surface tension $\sigma_0$ in 
$\S $\ref{32}. In this case, one recovers the correct equilibrium limit, and
order is predicted to disappear at $T=0$ for $p>p_{c}^{(\sigma _{0})}=(\sqrt{%
2}-1)^{2}\approx 0.1716$ (in agreement with MC results \cite{MarroDickman}).
However, the anomalous behavior of $\sigma _{0}$ that we describe in $\S $%
\ref{32} implies the emergence of an intermediate--$T$ region for $%
p_{c}<p<p_{c}^{\ast }\approx 0.18\,625$ where order sets in. This reentrant
behavior of $T_{c}(p),$ which is shown in Fig. \ref{pd}, is reminiscent of
the one in the phase diagram for systems subject to multiplicative noise, as
discussed below.

\section{Nonequilibrium Metastability and Nucleation\label{sect3}}

In order to produce metastability in the above nonequilibrium setting, we
initialize the system in a state with all spins up, $s_{i}=+1,i=1\ldots N.$
For small $h<0$ and $T<T_{c}(p),$ this configuration quickly relaxes to a
metastable state with magnetization $m=N^{-1}\sum_{i=1}^{N}s_{i}>0$. There
is, however, a tendency of the spins to line up in the direction of the
field, which competes with the tendency to maintain the local order induced
by the spins mutual interactions. The result for small $\vert h\vert$ is a metastable
state of long lifetime, i.e., the system fluctuates around the metastable
minimum for a long time, though eventually decays toward the truly stable
phase, which has $m<0$.

\subsection{The Metastable Phase: Mean--Field Theory (Pair Approximation)\label{31}}

The resulting metastable state is homogeneous at mesoscopic scales for times
shorter than its relaxation time, as expected \cite{Penrose1,Penrose2}, and
we can exploit this feature. In particular, we shall assume that: (\textit{i}%
) the system is completely homogeneous, and (\textit{ii}) it can be
described by the average density of up spins, $\rho (+)$, and by NN (pair)
correlations as captured by the density $\rho (s,s^{\prime })$ of NN spins.
Under these assumptions, all the spins will behave in the same manner, and
the evolution of a spin will depend exclusively on its current state and on
its four NNs. Denoting $x\equiv \rho (+)$ and $z\equiv \rho (+,+)$, we may
write $\rho (-)=1-x$, $\rho (+,-)=\rho (+)-\rho (+,+)=x-z$ and $\rho
(-,-)=1+z-2x$. In this approximation, the probability $Q(s,n)$ of finding a
spin in a state $s=\pm 1$ surrounded by $n\in \lbrack 0,4]$ up NN spins is 
\begin{equation}
Q(s,n)={\binom{4}{n}}\rho (s)\rho (+|s)^{n}\rho (-|s)^{4-n}={\binom{4}{n}}%
\frac{\rho (+,s)^{n}\rho (-,s)^{4-n}}{\rho (s)^{3}}\,,  \label{binomial}
\end{equation}%
where $\rho (\pm |s)=\rho (\pm ,s)/\rho (s)$ are marginal densities. The
pair of indexes $(s,n)$ defines the \emph{spin class} to which the given
spin belongs to. For periodic boundary conditions, there are 10 different
spin classes, as shown in Table I. The cost $\Delta \mathcal{H}(s,n)$ of
flipping any spin within a class is the same, so that the transition rate $%
\omega(\mathbf{s}\rightarrow \mathbf{s}^{i})\equiv \omega (s,n)$ in (\ref{rate}) only depends on $s$ and $n$, 
the spin at a node $i$ and the number of its up NN neighbors, respectively.

This information may be used to write down rate equations for the two
relevant observables in our mean field approximation, namely $x$ and $z.$ In
particular, the change in $x$ when a spin $s$ with $n$ up neighbors flips is 
$\delta x(s,n)=-s$, whereas the change in $z$ in the same process is $\delta
z(s,n)=-sn/4$. We thus obtain \cite{MarroDickman,Pablo1}:  
\begin{eqnarray}
\frac{\text{d}x}{\text{d}t} &=&G_{1}(x,z)\equiv \sum_{n=0}^{4}G\left(
x,z;n\right) \,,  \label{xzpair1} \\
\frac{\text{d}z}{\text{d}t} &=&G_{2}(x,z)\equiv \sum_{n=0}^{4}\frac{n}{4}%
\,G\left( x,z;n\right) \,,  \label{xzpair2}
\end{eqnarray}%
where 
\begin{equation}
G\left( x,z;n\right) \equiv {\binom{4}{n}}\left[ \frac{%
(x-z)^{n}(1+z-2x)^{4-n}}{(1-x)^{3}}\omega (-,n)\right. -\left. \frac{%
z^{n}(x-z)^{4-n}}{x^{3}}\omega (+,n)\right] .
\label{Gg}
\end{equation}%
\begin{table}[t]
\centerline{
\begin{tabular}{|c||c|c|c|}
\hline
Class & Central spin & Number of up neighbors & $\Delta {\cal H}$ \\
\hline \hline
1 & +1 & 4 & 8+2h \\
\hline
2 & +1 & 3 & 4+2h \\
\hline
3 & +1 & 2 & 2h \\
\hline
4 & +1 & 1 & -4+2h \\
\hline
5 & +1 & 0 & -8+2h \\
\hline \hline
6 & -1 & 4 & -8-2h \\
\hline
7 & -1 & 3 & -4-2h \\
\hline
8 & -1 & 2 & -2h \\
\hline
9 & -1 & 1 & 4-2h \\
\hline
10 & -1 & 0 & 8-2h \\
\hline \hline
\end{tabular}
}
\caption{{\protect\small Spin \textit{classes} for the two-dimensional Ising
model with periodic boundary conditions. The last column shows the energy
cost of flipping the central spin at each class.}}
\label{table}
\end{table}

\begin{figure}[t]
\centerline{
\psfig{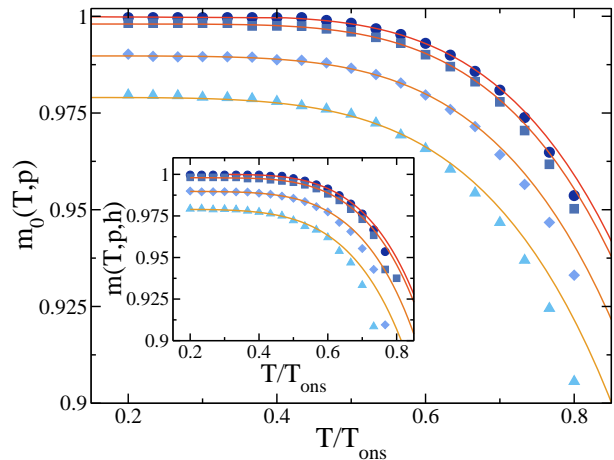}}
\caption{{\protect\small (Color) Temperature dependence of the steady--state
magnetization for $h=0$ and, from top to bottom, $p=0,$ $0.001,$ $0.005,$
and $0.01$. Lines are results from the mean--field (pair) approximation.
Inset: Magnetization in the metastable phase vs. $T$ for $h=-0.1$ and the
same same values of $p$ as in the main graph. }}
\label{magnet}
\end{figure}

Eqs. (\ref{xzpair1})-(\ref{Gg}) correspond to a mean-field 
Pair Approximation \cite{MarroDickman,Dickman}. This is a dynamic generalization 
of the first-order cluster variation method by Kikuchi \cite{Kikuchi}, and has been used 
to study other nonequilibrium systems with success \cite{Dickman}.
Our interest here is on the stationary solutions, $G_{1}(x_{\text{st}},z_{%
\text{st}})=G_{2}(x_{\text{st}},z_{\text{st}})=0.$ Both stable and
metastable solutions are locally stable against small perturbations, which
requires the associated Lyapunov exponents to be negative. This yields the
necessary and sufficient conditions 
\begin{gather}
\left( \partial _{x}G_{1}\right) _{\text{st}}+\left( \partial
_{z}G_{2}\right) _{\text{st}}<0\,,  \notag \\
\left( \partial _{x}G_{1}\right) _{\text{st}}\left( \partial
_{z}G_{2}\right) _{\text{st}}-\left( \partial _{z}G_{1}\right) _{\text{st}%
}\left( \partial _{x}G_{2}\right) _{\text{st}}>0\,.  \label{Hurwitz}
\end{gather}%
On the other hand, the condition $(\partial _{x}G_{1})_{\text{st}}=0$
signals for $h=0$ an incipient or marginal instability corresponding to an
order--disorder phase transition with critical point $(x_{\text{st}}^{\text{c%
}},z_{\text{st}}^{\text{c}})=(1/2,1/3)$. It follows that 
\begin{equation}
T_{c}(p)=-4\left[ \ln \left( \frac{3}{4}\sqrt{\frac{1-4p}{1-p}}-\frac{1}{2}%
\right) \right] ^{-1}\,.  \label{TcMF}
\end{equation}%
This function is depicted in Fig. \ref{pd} (see the related discussion above
in $\S $\ref{model}). The stationary state $(x_{\text{st}},z_{\text{st}})$
may be obtained numerically from the nonlinear equations $G_{1}(x_{\text{st}%
},z_{\text{st}})=G_{2}(x_{\text{st}},z_{\text{st}})=0$ subject to the local
stability conditions in eq. (\ref{Hurwitz}). For $h=0$, the up--down symmetry leads to pairs of
locally--stable steady solutions, namely $(x_{\text{st}},z_{\text{st}})$ and 
$(1-x_{\text{st}},1+z_{\text{st}}-2x_{\text{st}})$. The result is
illustrated in Fig. \ref{magnet}. This also illustrates the expected
agreement with Monte Carlo results at low and intermediate temperature for
any $p$. In particular, this agreement is good for temperatures below $0.75 \,T_c(p)$. 
The fact that increasing $p$ at fixed $T$ decreases the
magnetization reveals that the nonequilibrium perturbation tends to increase
disorder, as expected. In this way we can identify the parameter $p$, which weights the competition between 
different heat baths leading to nonequilibrium behavior, as a source of \emph{nonequilibrium noise}, 
as compared with thermal fluctuations induced by $T$. 
For small enough fields, the situation closely resembles the case $%
h=0$, but the up--down symmetry is now broken, and locally--stable steady
states with magnetization opposite to the applied field are metastable, see
inset to Fig. \ref{magnet}.

\subsection{Nonequilibrium Surface Tension\label{32}}

\begin{figure}
\centerline{
\psfig{file=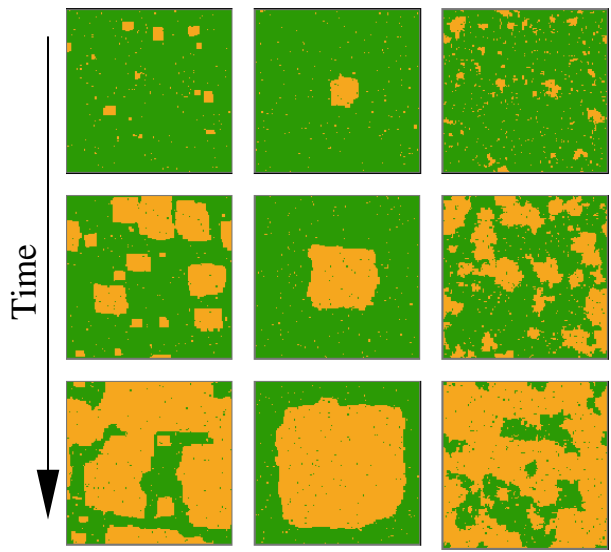,width=8cm}}
\caption{{\small (Color) Configurations as the system escapes from the metastable 
state, at different times as indicated, for $L=128,$ $p=0.01,$ $h=-0.25,$ and, from 
left to right, $T=0.1,$ 0.3 and 0.7 in units of $T_{\text{ons}}.$}}
\label{snap}
\end{figure}

The above mean--field approximation neglects fluctuations, so that it cannot
account for the relaxation of the metastable phase. One expects that, for long enough
times, fluctuations will allow for a decay toward the stable phase. Direct
inspection of escape configurations as the ones in Fig. \ref{snap} indicates
that the metastable--stable transition is a highly inhomogeneous process
triggered by large stable--phase clusters, and a detailed comparison (not
shown) suggests that the nonequilibrium and equilibrium cases are
characterized by the same type of relevant excitations. These grow or shrink
into the metastable sea depending upon the competition between their
surface, which hampers cluster growth, and their bulk, which favors it.

The nucleation and growth of a cluster is controlled in equilibrium, $p=0,$ by the
zero--field interfacial free--energy, or surface tension. For $p>0,$ even
though a proper bulk free--energy function does not exist, it seems sensible
to assume that the properties of the nonequilibrium interface will be
determined by a sort of nonequilibrium or effective \emph{surface tension} $%
\sigma _{0}(T,p)$ \footnote{In general, the surface tension depends on the orientation angle 
$\phi$ of the interface with respect to one of the lattice axes, $\sigma_0(\phi,T,p)$. However, 
the surface tension entering nucleation theory is the one defined along a primitive lattice 
vector \cite{Gunton1,Langer1,Rikvold,Debenedetti}, so $\phi=0$ henceforth.}.
This relies on the assumption that the normalization of the probability measure
for interface configurations can be interpreted as a nonequilibrium 
\emph{partition function}. More formally, let $\mathbf{y}$ denote an interface
microscopic configuration ---e.g., the set of heights $y_{k}$, $k=1\ldots L$%
, that characterize the steps of the interface---, and $Z^{-1}\mu (\mathbf{y}%
;T,p)$ the associated invariant measure, which for $0<p<1$ is non--Gibbsian. Then, we define 
$\sigma_{0}(T,p)=-(\beta L)^{-1}\ln Z$ \cite{Pablo2}. Similar hypotheses have already been tested
in relation with other nonequilibrium phenomena. For instance, concerning
nonequilibrium phase transitions, the distribution of complex zeros of the
invariant measure normalization factor has recently been shown to obey the
Lee--Yang picture in the case of the asymmetric simple exclusion process 
\cite{LY1} and for systems with absorbing states \cite{LY2}; see also \cite{Pablo1,Pablo2}.
Interestingly, the normalization $Z$ is defined in general up to a multiplicative factor \cite{LY1}.
This is related to the fact that one has to arbitrarily 
choose a reference interface configuration from which to measure the relative 
probabilities of all other microscopic configurations. 
However, in the case of our model, the ambiguity in $Z$ can be easily resolved by noting that a well-defined 
equilibrium limit exists, $p=0$. In this way, demanding $Z$ to converge to the right equilibrium limit 
for $p=0$, the ambiguity disappears. In more general nonequilibrium cases, where no proper 
equilibrium limit can be defined, it is expected that the spurious multiplicative factor
won't play any relevant physical role \cite{LY1}. 

\begin{figure}[t]
\centerline{
\psfig{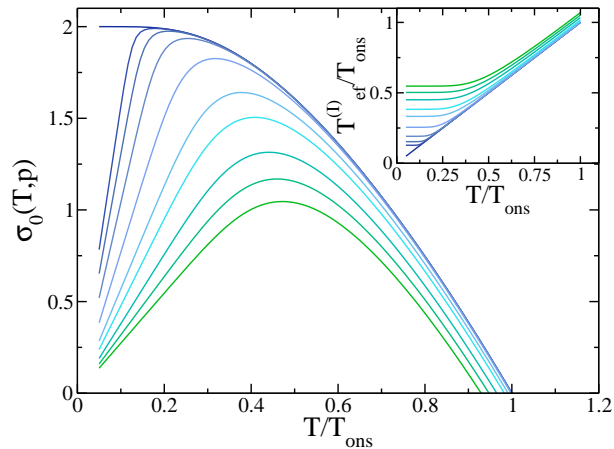}}
\caption{{\protect\small (Color) The nonequilibrium \emph{surface tension}
as a function of $T$ for, from top to bottom, $%
p=0,10^{-6},10^{-5},10^{-4},10^{-3},5\times 10^{-3},10^{-2},2\times
10^{-2},3\times 10^{-2},$ and $4\times 10^{-2}$. For any $p>0,$ the surface
tension behaves non-monotonously, contrary to the equilibrium case. Inset:
The effective interface temperature $T_{\text{ef}}^{\text{(I)}}(T,p)$, as
defined in the main text, versus $T$ for the same values of p than in the
main graph. Notice that $T_{\text{ef}}^{\text{(I)}}(T,p>0)$ strongly
deviates from $T$ in the low-temperature regime. }}
\label{sigma}
\end{figure}

Computing analytically the exact interfacial invariant measure $Z^{-1}\mu $
is beyond our goals, but a solid--on--solid approximation \cite{SOS} will
suffice. It thus follows (we refer the reader to Ref. \cite{Pablo2} for
further details) that $\sigma _{0}(T,p)$ has non--monotonous $T$ dependence
for any $p>0$ and a maximum at a non--trivial value of $T$ which depends on $%
p$, see Fig. \ref{sigma}. This peculiar behavior can be understood in simple
terms. From low-- and high--$T$ expansions of $\sigma _{0}(T,p),$ one
obtains that 
\begin{equation}
\sigma _{0}\approx \frac{T}{T_{\text{ef}}^{\text{(I)}}}\,\sigma _{\text{e}%
}\,+\,\text{small corrections}\,,  \label{sigexp}
\end{equation}%
where $T_{\text{ef}}^{\text{(I)}}(T,p)$ is an effective \textit{interfacial
temperature} and $\sigma _{\text{e}}(T)$ is the exact equilibrium surface tension 
\cite{exact}. At high$-T$, where thermal fluctuations dominate over the
nonequilibrium perturbation, $T_{\text{ef}}^{\text{(I)}}$ is proportional to 
$T$ ---see inset to Fig. \ref{sigma}---, so that $\sigma _{0}$ is
proportional to $\sigma _{\text{e}}$ which decreases with $T.$ However, $T_{%
\text{ef}}^{\text{(I)}}$ saturates to a constant as $T\rightarrow 0,$
namely, $\lim_{T\rightarrow 0}T_{\text{ef}}^{\text{(I)}}(T,p)=2/\ln [(1-%
\sqrt{p})/(p+\sqrt{p})]$ \cite{Pablo2}. This, which is due to non--vanishing
nonequilibrium fluctuations, implies that $\sigma _{0}$ is a
linearly--increasing function of $T$ at low--$T$. Consequently, it follows a
non--monotonous $T$--dependence of $\sigma _{0}$ with a maximum which
roughly agrees with the crossover observed in $T_{\text{ef}}^{\text{(I)}}$.

The properties of the nonequilibrium interface may be obtained from the
resulting $\sigma _{0}(T,p).$ For instance, the temperature dependence of
this function leads to the phase diagram $T_{c}(p).$ In equilibrium, the
interface free energy goes to zero as $T$ increases toward $T_{\text{ons}},$
and there is no surface tension in the disordered phase for $T>T_{\text{ons}%
} $ \cite{Hartmann}. Therefore, one may identify $T_{c}(p)$ as the (finite)
temperature for which $\sigma _{0}(T,p)=0.$ We thus obtain the result in
Fig. \ref{pd}. On the other hand, $\sigma _{0}$ is linear in $T$ as 
$T\rightarrow 0$ (see Fig. \ref{sigma}), namely $\sigma _{0}(T,p)\approx \alpha (p)T$, with 
\cite{Pablo2}  
\begin{equation}
\alpha(p)=\ln \left( \frac{1-\sqrt{p}}{p+\sqrt{p}} \right) \ .
\label{alfap}
\end{equation}
The condition $\alpha(p_{c})=0$ thus signals the onset of disorder at low temperature. 
This yields $p_{c}=(\sqrt{2}-1)^{2}\approx 0.1716$, in excellent agreement with
previous MC simulations \cite{MarroDickman,Pedro1}.

\subsection{Effective Free--Energy and Nucleation\label{33}}

The structural similarities between equilibrium and nonequilibrium
excitations responsible for the metastable--stable
transition allows one to write an effective \emph{free energy}
associated to a cluster of radius $R.$ In formal analogy with the
equilibrium case \cite{Rikvold}, we assume that 
\begin{equation}
\mathcal{F}(R)=\gamma \left[ 2\Omega R\sigma _{0}-\Omega R^{2}2m_{0}|h|%
\right] \,.  \label{free}
\end{equation}%
Here, $\sigma _{0}$ is the nonequilibrium surface tension in the previous
subsection, $m_{0}(T,p)$ is the zero--field spontaneous magnetization, which
follows from the mean--field scheme in $\S $\ref{31}, and $\Omega (T,p)$ is
the cluster form factor, which relates the cluster linear length $R$ to its
volume, $C=\Omega R^{2}.$ This may be computed from $\sigma _{0}$ via the
Wulff construction \cite{Wulff,Pablo2}. The multiplicative factor in eq. (\ref{free}), 
$\gamma \simeq 1,$ is a phenomenological parameter, which might
have a very weak dependence on $T$ and $p,$ intended to capture small
corrections to classical nucleation theory
due to correlations between different clusters, inhomogeneous magnetization 
profiles within a cluster, etc \cite{Debenedetti}.

\begin{figure}[t]
\centerline{
\psfig{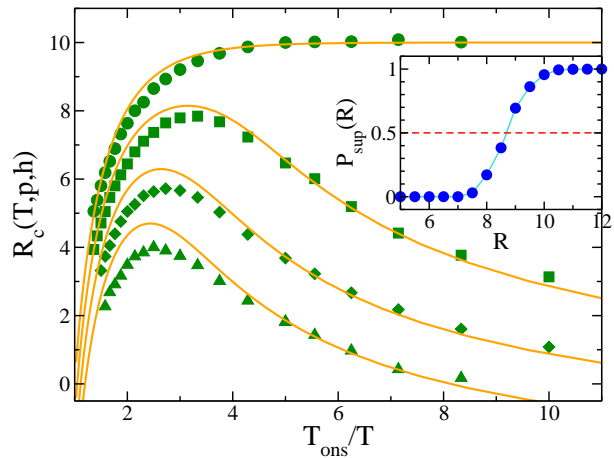}}
\caption{{\protect\small (Color) $\mathcal{R}_{c}$ vs. $T^{-1}$ for $L=53$, $%
h=-0.1$ and, from top to bottom, $p=0$, $0.001$, $0.005$ and $0.01$. Symbols
are MC results. Lines correspond to the theoretical prediction. The $n^{th}$
curve (from bottom to top) has been shifted $(4-n)$ units in the $\hat{y}$%
-axis. Inset: MC results for $P_{\text{spc}}(R)$ as a function of $R$ for $%
L=53$ at $T=0.4\,T_{\text{ons}}$, $p=0$, and $h=-0.1$. All the data in these
graphs correspond to an average over $1000$ independent experiments.}}
\label{Rc}
\end{figure}

The first term in $\mathcal{F}(R)$ measures the cost of the cluster
interface, while the second term stands for the bulk gain. In this way,
small clusters ---with a large surface/volume ratio--- tend to shrink, while
large clusters tend to grow. The critical cluster radius, $\mathcal{R}_{c},$
which corresponds to the maximum of $\mathcal{F}(R),$ separates these two
tendencies. One has that%
\begin{equation}
\mathcal{R}_{c}=\frac{\sigma _{0}}{2m_{0}|h|}\,.  \label{gotacrit}
\end{equation}%
We measured this quantity in MC simulations. With this aim, one produces an
initial state with a single \emph{square} cluster of down spins and side $2R$
(the stable phase) in a sea of up spins. This is highly unstable, so that
any \textit{subcritical} initial cluster, $R<\mathcal{R}_{c}(T,p),$ will
very quickly shrink, while \textit{supercritical} ones, $R>\mathcal{R}%
_{c}(T,p),$ will rapidly grow to cover the whole system. Since our dynamics
is stochastic, we define the probability that a cluster of size $R$ is
supercritical, $P_{\text{spc}}(R)$. This is measured in practice by simply
repeating many times the simulation and counting the number of times that
the initial cluster grows to cover the system. $\mathcal{R}_{c}$ is then
defined as the solution of $P_{\text{spc}}(\mathcal{R}_{c})=0.5.$ As
illustrated in the inset to Fig. \ref{Rc}, $P_{\text{spc}}(R)$ shows a sharp
transition from $0$ to $1$, which allows for a relatively accurate estimate
of $\mathcal{R}_{c}$. \footnote{Our numerical results on ${\cal R}_c$ depend 
in principle on the shape of the initial excitation, which we choose squared. 
However, we checked that using spherical initial excitations (good at high $T$) one 
obtains estimates for ${\cal R}_c$ which are very close to those obtained with 
squared excitations (expected at low $T$, see Ref. \cite{Pablo2}).}
The agreement shown in Fig. \ref{Rc} between these
results and our analytical predictions is excellent for temperatures well
below $T_{c}(p)$. Interesting enough, $\mathcal{R}_{c}(T,p>0,h)$ exhibits
non--monotonous $T$--dependence, as expected from the anomalous low--$T$
behavior of $\sigma _{0}$.

\begin{figure}[t]
\centerline{
\psfig{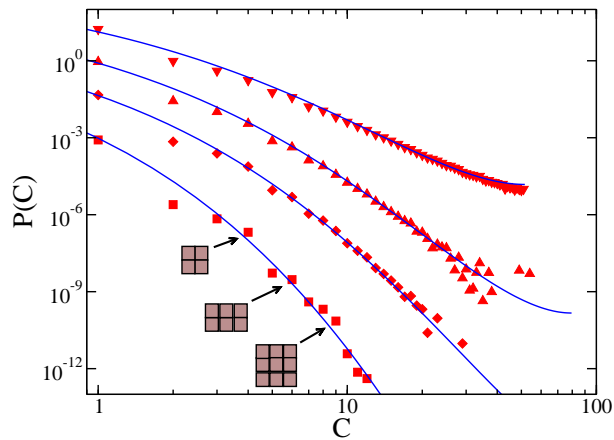}}
\caption{{\protect\small (Color) Cluster distribution $P(C)$ for $T=0.2\,T_{%
\text{ons}}$, $L=53$, $h=-0.1$ and, from bottom to top, $p=0.001$, $0.005$, $%
0.01$, $0.02$, with $\protect\gamma =0.815$, $0.82$, $0.83$, $0.85$,
respectively. Lines are theoretical predictions and points are MC results.
The $n^{th}$ curve (from bottom to top) has been rescaled by a factor $%
10^{(n-1)}$ in the $\hat{y}$-axis. }}
\label{clust}
\end{figure}

Our ansatz in eq. (\ref{free}) also implies that the fraction of stable--phase clusters of volume $%
C=\Omega R^{2}$ in the metastable phase follows a Boltzmann distribution
\begin{equation}
P(C)=\mathcal{M}^{-1}\exp \left[ -2\beta \gamma \left( \sigma _{0}\sqrt{%
\Omega C}-m_{0}|h|C\right) \right]   \label{distrib}
\end{equation}%
Here, the normalization $\mathcal{M}=2\Theta /(1-m)$, where $\Theta
=\sum_{C=1}^{C_{\ast }}C\exp [-\beta \mathcal{F}(C)]$ with $C_{\ast }=\Omega 
\mathcal{R}_{c}^{2}$, is defined so that the metastable state has
magnetization $m(T,p,h)$ as derived in $\S $\ref{31}. The cluster
distribution $P(C)$ which is obtained in MC simulations for times much
shorter than the relaxation time is shown in Fig. \ref{clust}. This compares
very well with the analytical prediction, eq. (\ref{distrib}). For $%
p=0.001$, we observe a non--trivial structure in $P(C)$ which is not
captured by our continuous description. This is due to the underlying
lattice anisotropy, which for low--$T$ and small $p$ gives rise to
resonances in $P(C)$ for clusters with complete \emph{shells}, i.e. $2\times
2$, $3\times 2$, $3\times 3;$ see Fig. \ref{clust}. For larger $p$ and/or $T$%
, fluctuations smear out these resonances. Notice in Fig. \ref{clust} that
the nonequilibrium perturbation $p$ enhances fluctuations and favors larger
clusters. Similar good agreement for $P(C)$ is found in a wide range of
values for $T$, $p$ and $h$. On the other hand, systematic corrections
in $P(C)$ show up for microscopic clusters. These appear because our 
main ansatz, eq. (\ref{free}), is based on mesoscopic considerations and 
therefore is only valid for large enough clusters. 
The observed deviations won't play however a relevant role in our discussion, as far as 
the critical cluster remains large enough, as it is the case here, see Fig. \ref{Rc}.

\begin{figure}[t]
\centerline{
\psfig{file=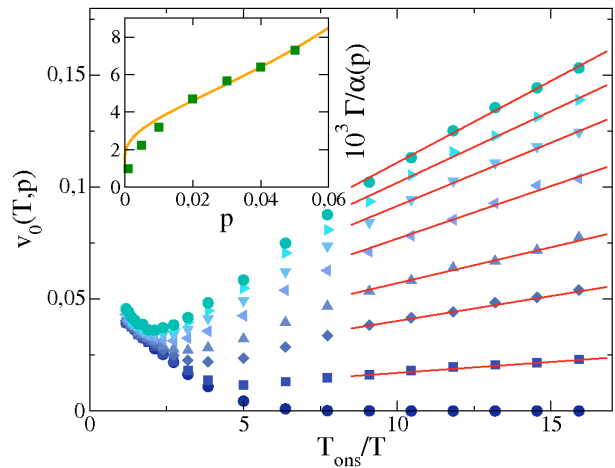,width=8cm}}
\caption{{\small (Color) Propagation velocity $v_0$ vs. $T^{-1}$ for $h=-0.1$ and
$p=0, \ 0.001,  \ 0.005, \ 0.01, \ 0.02, \ 0.03, \ 0.04, \ 0.05$ (from bottom to top). Curves are 
linear fits to data. Inset: Slope of linear fits vs. $p$. The line is the theoretical prediction.}}
\label{vel}
\end{figure}

The assumption above on $\mathcal{F}(R)$ implies a force per unit area which
controls the growth of supercritical clusters. It follows that the
propagation velocity of a cluster in the large--size limit will obey the
Allen--Cahn expression \cite{Gunton1,Langer1}:%
\begin{equation}
v_{0}=\frac{2\Gamma ^{\prime }m_{0}|h|}{\sigma _{0}},  \label{vel0}
\end{equation}%
where $\Gamma ^{\prime }$ is a non--universal constant. We measured $v_{0}$
in MC simulations of an \emph{infinitely--large} cluster. This was
implemented in practice by a flat interface separating the stable phase from
the metastable one which constantly propagates. The results, which are shown
in Fig. \ref{vel}, agree with our predictions. We also observe that $v_{0}$
exhibits non--monotonous $T$--dependence, as expected given the behavior
described above for $\sigma _{0}$. This implies, in particular, that, at low 
$T,$ cooling the system favors the interface propagation or, in other words,
that the interface dynamics becomes more sluggish as we raise the
temperature in the low--$T$ regime.

As discussed before, $\sigma _{0}$ increases linearly with $T$ at low
temperature, namely, $\sigma _{0}\approx \alpha (p)T$ as $T\rightarrow 0$
with $\alpha (p)$ given in eq. (\ref{alfap}). Therefore we expect in this regime that $%
v_{0}\approx \left[ \Gamma \diagup \alpha \left( p\right) \right] \left( T_{%
\text{ons}}\diagup T\right) $ with $\Gamma =2\Gamma ^{\prime }|h|T_{\text{ons%
}}^{-1}$ assuming that $m_{0}(T\rightarrow 0,p)\sim \mathcal{O}(1)$. This
low--$T$ behavior, linear in $T^{-1}$, is also confirmed by simulations, see the inset to
Fig. \ref{vel}, with $\Gamma \approx 0.0077$.

The nucleation rate $\mathcal{I}$, i.e., the probability that a critical
cluster nucleates per unit time and per unit volume, may be written as%
\begin{equation}
\mathcal{I}=A|h|^{\delta }\exp [-\beta \mathcal{F}(\mathcal{R}_{c})]\,,
\label{nucl}
\end{equation}%
where $A(p)$ is a non--universal amplitude and $\delta \approx 3$ for random
updatings \cite{Rikvold}. Two main relaxation patterns arise depending on the
interplay between the relevant length scales in the problem. In fact, one
needs to deal with the competition between the system size, $L,$ and the 
mean cluster separation, $\mathcal{R}_{0}(T,p,h)$. The two other important length 
scales, namely the critical radius, $\mathcal{R}_{c}$, and the thermal correlation length 
in the metastable phase, $\xi_{\text{ms}}$, do not compete with the former two given that 
they are much smaller in the most interesting regime, 
$\xi_{\text{ms}} \ll \mathcal{R}_{c} \ll \mathcal{R}_{0}, L$. Now $\mathcal{R}%
_{0}$ can be calculated together with its associated time scale $t_{0}$ by
requiring that $\mathcal{R}_{0}=v_{0}t_{0},$ which states that a
supercritical cluster propagating with velocity $v_{0}$ will grow a
distance $\mathcal{R}_{0}$ in a time $t_{0},$ and that $\mathcal{R}%
_{0}^{2}t_{0}\mathcal{I}=1,$ which states that the probability of nucleating
a volume $\mathcal{R}_{0}^{2}$ in a time $t_{0}$ is (by definition) one.
Therefore, $\mathcal{R}_{0}=(v_{0}/\mathcal{I})^{1/3}$. For $\mathcal{R}%
_{0}\gg L$, the random nucleation of a \emph{single} critical cluster is the
relevant excitation. In this regime, known as \emph{Single--Droplet} (SD)
region, the lifetime of the metastable state is $\tau _{\text{SD}}=(L^{2}%
\mathcal{I})^{-1}$. On the other hand, for $\mathcal{R}_{0}\ll L,$ the
metastable--stable transition proceeds via the nucleation of many critical
clusters, a regime known as \textit{Multi--Droplet} (MD) region, and $\tau _{%
\text{MD}}=[\Omega v_{0}^{2}\mathcal{I}/(3\ln 2)]^{-1/3}$ \cite{Rikvold}.
Summing up,
\begin{equation}
\tau (T,p,h)\propto \left\{ 
\begin{array}{ll}
\left(L^{2}\mathcal{I}\right)^{-1} & \mathcal{R}_{0}\gg L\,\text{(SD)}\,, \\ 
\left( v_{0}^{2}\mathcal{I}\right) ^{-1/3} & \mathcal{R%
}_{0}\ll L\,\text{(MD)}\,.%
\end{array}%
\right. 
\end{equation}%
The crossover between these regimes, given by the condition $\mathcal{R}%
_{0}=L$, characterizes the dynamic spinodal point, namely, 
\begin{equation}
|h_{\text{DSP}}|(T,p)=\frac{\Omega \gamma \sigma _{0}^{2}}{6m_{0}T\ln L}.
\label{hdsp}
\end{equation}%
That is, one should observe the SD regime for $|h|<|h_{\text{DSP}}|,$ while
the MD regime is expected to dominate otherwise. 
\begin{figure}[t]
\centerline{
\psfig{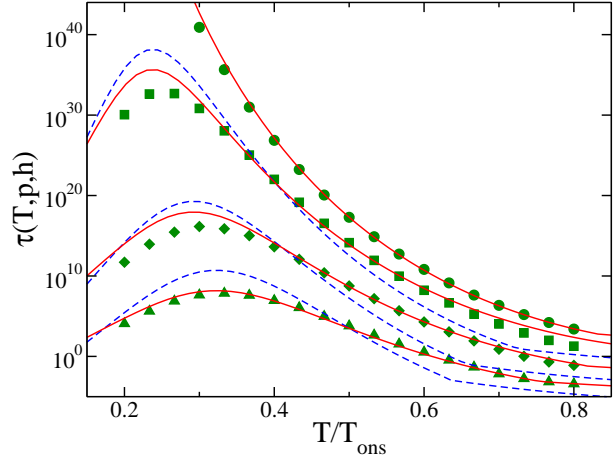}}
\caption{{\protect\small (Color) Lifetime $\protect\tau $ of the metastable
state vs. $T$ for the same values of $p$ and conditions than in Fig. \protect
\ref{Rc}. The $n^{th}$ curve (from top to bottom) is rescaled by a factor $%
10^{-2(n-1)}$. Solid lines are theoretical predictions for (from top to
bottom) $\protect\gamma =1,\ 0.85,\ 0.77,\ 0.65$. Amplitudes $A(p)$ are in
the range $[10^{-3},10^{-2}]$. Dashed lines are predictions for $\protect%
\gamma =1$. }}
\label{vida}
\end{figure}

We estimated $\tau $ from MC simulations by defining it as an average of the
first-passage time to zero magnetization. Due to the strong local
stability that characterizes the metastable state at low temperature, $\tau $
may be extremely large in practice. (For example, this goes up to $10^{40}$
MC step per spin in Fig. \ref{vida}. Assuming the MC time unit corresponds to the
inverse of the typical phonon frequency\footnote{Certain rare-earth magnetic materials 
can be modeled to a good approximation by the Ising Hamiltonian \cite{Wolf}. 
When we talk here about the phonon frequency we refer to the phonon frequency 
in one of these real magnetic system with Ising-like properties.}, which is of order of $10^{-12}$ s,
this corresponds to $\tau \sim 10^{28}$ s!). Therefore we needed to use an
efficient rejection--free algorithm. Our choice was the $s-1$ variant of
the MC algorithm with absorbing Markov chains, together with the
slow--forcing approximation \cite{MCAMC}. Fig. \ref{vida} shows our results
for $\tau $ vs. $T$ and different values of $p$, together with theoretical
predictions. Interesting enough, we observe that $\tau $ increases with $T$ at 
low--$T$ for fixed $p>0.$ That is, the local stability of \emph{nonequilibrium} 
metastable states is enhanced at low--$T$ as the strength of the thermal noise ($T$) increases. 
This, which is in contrast with the Arrhenius curve observed
in equilibrium, closely resembles the noise--enhanced stability (NES)
phenomenon reported in experiments on unstable systems \cite{Mantegna}. On
the other hand, increasing $p$ for fixed $T$ always results in shorter $\tau$. 
This complex phenomenology is captured by our simple ansatz eq. (\ref{free}), 
which traces back the NES phenomenon to the low--$T$ anomaly in $\sigma _{0},$
which in turn is a nonlinear cooperative effect due to the interplay between
thermal and nonequilibrium fluctuations. That is, though any of these two 
\textit{noise sources} ($T$ and $p$) will separately induce disorder, their combined effect
results in (resonant) stabilization of the nonequilibrium metastable state.

\begin{figure}[t]
\centerline{
\psfig{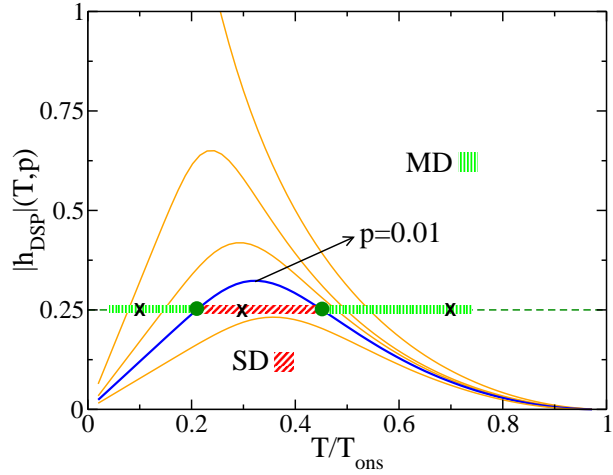}}
\caption{{\small (Color) The dynamic spinodal $|h_{\text{DSP}}|,$ as defined
in eq. (\protect\ref{hdsp}), versus $T$ for $p=0,$ $0.001,$ $0.005,$ $0.01$,
and 0.02 from top to bottom, respectively. We explicitly indicate the $T$
ranges corresponding to the SD and MD regimes in the case of the example in
Fig. \protect\ref{snap}, namely, for $|h|=0.25$ and $p=0.01.$ The symbols $\times$ 
mark the temperatures $T/T_{\text{ons}}$ = 0.1, 0.3 and 0.7 for $|h|=0.25.$ 
Notice that, unlike at equilibrium, there is here a low-$T$ regime where 
the MD scenario holds. This is confirmed in Fig. \protect\ref{snap}.}}
\label{DSP}
\end{figure}

This interesting effect is also evident in the morphology of the
metastable--stable transition. In particular, $|h_{\text{DSP}}|$ inherits
the non--monotonous $T$--dependence of $\sigma _{0}$, see Fig. \ref{DSP},
resulting in a novel MD regime at low--$T$ not observed in equilibrium. For
instance, for $h=-0.25$ and $p=0.01$, we expect from Fig. \ref{DSP} MD
morphology for, e.g., $T=0.1\,T_{\text{ons}}$ and $T=0.7\,T_{\text{ons}}$
but SD behavior for $T=0.3\,T_{\text{ons}},$ which is confirmed by direct
inspection of escape configurations such as the ones in Fig. \ref{snap}.
Notice that the SD decay mechanism is a finite--size effect, and that MD
will be the only relevant mechanism for large enough $L.$ However, the
relevance of the SD regime vanishes \emph{logarithmically} with $L$ ---see
eq. (\ref{hdsp})--- so that it could be observable in mesoscopic samples, at
least \cite{Rikvold}.

\subsection{Langevin Type Description\label{34}}

We may get further insight into the above stochastic resonance phenomena and
the peculiar low--$T$ behavior by rewriting the rate (\ref{rate}) as 
\begin{equation}
p+(1-p)\Psi (\beta \Delta \mathcal{H}_{i})\equiv \Psi (\beta _{\text{ef}%
}\Delta \mathcal{H}_{i})\,.  \label{rateff}
\end{equation}%
The resulting effective temperature $T_{\text{ef}}\equiv \beta _{\text{ef}%
}^{-1}$ then follows as
\begin{equation}
T_{\rm{ef}}(T,p;\Delta {\cal H}_i)= \frac{T \Delta {\cal H}_i}{\Delta {\cal H}_i +
T \ln \left(\frac{\displaystyle 1-p}{\displaystyle 1+p \textrm{e}^{\beta \Delta {\cal H}_i}} \right)}
\label{TeffG}
\end{equation}
For any $p>0$, $T_{\rm{ef}}$ changes from spin to spin and depends on the 
\emph{local order} as characterized, for instance, by the number $\eta_i$ of
neighbors that are in the same state than the central spin (in fact, one can write 
$\Delta {\cal H}_i=4(\eta_i-2)+2s_ih$, see Table I): the larger $\eta_i$, the higher the 
amount of local order, and the larger the local effective temperature. This observation is 
not restricted to the Glauber function we use for $\Psi$ in this paper, see $\S$II and 
eq. (\ref{rate}), but holds for many other known forms for $\Psi$. For instance, for the 
Metropolis case, which is analytically simpler for the purpose, one obtains for $h\rightarrow 0$ that 
\begin{equation}
T_{\text{ef}}^{(\eta_i)}(T,p)=\left\{ 
\begin{array}{ll}
-8\left( \ln \left[ p+(1-p)\text{e}^{-8\beta }\right] \right) ^{-1} & \eta_i=4\,, \\ 
-4\left( \ln \left[ p+(1-p)\text{e}^{-4\beta }\right] \right) ^{-1} & \eta_i=3\,, \\ 
T & \eta_i =2,1,0\,,%
\end{array}%
\right.   \label{TeffM}
\end{equation}%
such that $T_{\text{ef}}^{\text{(4)}}\geq T_{\text{ef}}^{\text{(3)}}\geq T_{%
\text{ef}}^{\text{(2,1,0)}}>0$, the equality holding only in the equilibrium
limit $p=0$. Therefore, in all cases, the strength of fluctuations that affect a spin
increases with the local order parameter. This is the fingerprint of 
\emph{multiplicative noise,} which suggest us to write a Langevin equation 
with the hope it will capture the physics of the problem. Defining the coarse-grained 
order parameter as the local magnetization field $\psi({\bf r},t)$, we propose the following Langevin equation
\begin{equation}
\frac{\partial \psi({\bf r},t) }{\partial t}=D \nabla^2 \psi({\bf r},t) + \psi({\bf r},t) -\psi({\bf r},t) ^{3}+h+\sqrt{D+\mu \psi({\bf r},t)^{2}} \xi ({\bf r},t),  
\label{Langevin1}
\end{equation}
where $\xi ({\bf r},t)$ is a Gaussian white noise with $\langle \xi ({\bf r},t)\rangle =0$
and $\langle \xi ({\bf r},t)\xi ({\bf r}^{\prime},t^{\prime })\rangle =
2\delta (t-t^{\prime })\delta ({\bf r}-{\bf r}^{\prime })$, $D$
is the strength of the thermal (\emph{additive}) noise, $h$ is an external field
parameter, and $\mu $ is the renormalized version of the nonequilibrium
parameter $p$. This equation describes the coarse-grained dynamics of a Ising-like 
system in the broken-symmetry phase under an external magnetic field, with a noise term 
whose amplitude increases as the local order parameter $\psi({\bf r},t)$ increases, in close 
analogy with our observations above. 
Solving the full spatially-dependent problem, eq. (\ref{Langevin1}), is a complex endeavor 
which we will not undertake here. However, as a proof of concept, one
can analytically solve a simplified model corresponding to the mean-field, 0-dimensional form
of the above Langevin equation:
\begin{equation}
\frac{\partial \psi }{\partial t}=\psi -\psi ^{3}+h+\sqrt{D+\mu \psi ^{2}} \xi (t) \, ,
\label{Langevin}
\end{equation}
where now $\psi(t)$ has no spatial dependence. 
Despite the obvious simplifications, this equation still contains the essential
competition between thermal ($D$) and nonequilibrium ($\mu$) fluctuations
in a metastable potential that characterizes our system\footnote{As a result of the 
homogeneity hypothesis associated to the mean-field approximation, eq. (\ref{Langevin}) 
has no spatial structure and therefore is unable to describe the compact excitations 
responsible of the metastable state decay in the microscopic model. Despite this obvious 
difficulty, eq. (\ref{Langevin}) still contains the relevant competition between additive and 
multiplicative fluctuations which gives rise to resonant behavior in this model.}. In particular, 
eq. (\ref{Langevin}) describes the motion of a \emph{Brownian particle} in
an asymmetric bimodal potential, namely, $V(\psi )=-\frac{1}{2}\psi ^{2}+%
\frac{1}{4}\psi ^{4}-h\psi $ ---which is depicted in the inset to Fig. \ref%
{vidafi}. The particle is subject to fluctuations which increase with $\psi^{2}$
and whose amplitude remains non--zero as $D\rightarrow 0$ for any $\mu >0$. 
Notice that the \emph{noise strength} $\sqrt{D+\mu \psi ^{2}}$ increases
with the order parameter ($\sim\psi ^{2}$), in accordance with our observation
above concerning eqs. (\ref{TeffG})-(\ref{TeffM}). For $\mu =0,$ this reduces to the
additive theory which describes simple symmetry--breaking systems \cite{vanK}
while, for $\mu >0,$ additive and multiplicative fluctuations compete in eq.
(\ref{Langevin}).

\begin{figure}[t]
\centerline{
\psfig{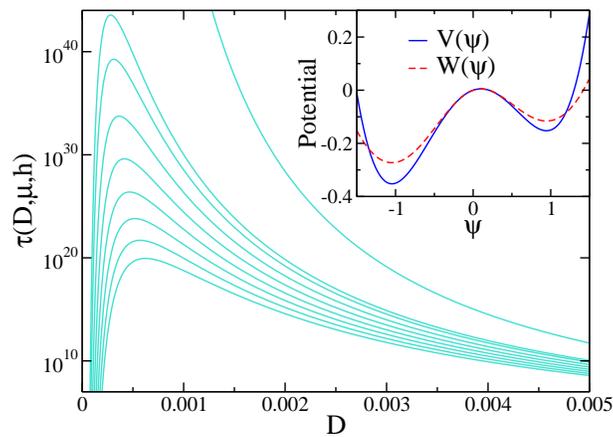}}
\caption{(Color) Lifetime $\tau$, as derived from 
the Langevin equation (\ref{Langevin}), versus $D$ for $h=-0.1$ and
for $\mu \in [0,6\times 10^{-3}]$ increasing from top to bottom. The inset 
illustrates the potential $V$ and the effective potential $W$ as defined 
in the main text. 
}
\label{vidafi}
\end{figure}

\begin{figure}[t]
\centerline{\psfig{file=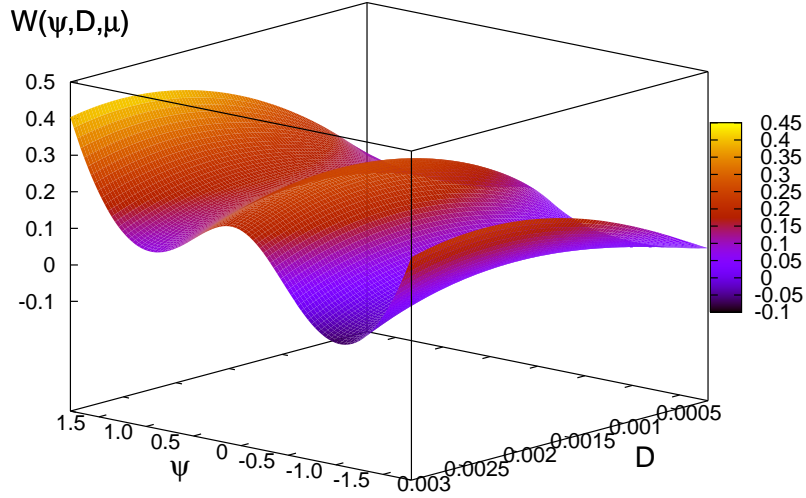,width=8cm,angle=-90}}
\caption{{\protect\small (Color) Effective potential $W(\psi,D,\mu)$ for 
$\mu=4\times 10^{-3}$ and $h=-0.1$. Note how the energy barrier between 
the two local minima changes as a function of $D$ for this $\mu>0$.}}
\label{wpot}
\end{figure}

The Stratonovich stochastically--equivalent Fokker--Planck equation of eq. (%
\ref{Langevin}) is $\partial _{t}P(\psi ,t)=-\partial _{\psi }[A(\psi
)P(\psi ,t)]+\frac{1}{2}\partial _{\psi }^{2}[B(\psi )P(\psi ,t)]$ with $%
A(\psi )=(1+\mu )\psi -\psi ^{3}+h$ and $B(\psi )=2(D+\mu \psi ^{2})$. This
yields the steady distribution 
\begin{equation}
P_{\text{st}}(\psi )=\frac{\Lambda }{2\sqrt{\mu (D+\mu \psi ^{2})}}\,\text{%
exp}\left[ -\frac{W(\psi )}{D}\right] ,  \label{steadydist}
\end{equation}%
where $\Lambda $ is a normalization constant, and 
\begin{equation}
W(\psi )=\frac{D}{\mu }\left[ \frac{1}{2}\psi ^{2}-\frac{1}{2}\left( 1+\frac{%
D}{\mu }\right) \ln \left( \frac{D}{\mu }+\psi ^{2}\right) -h\sqrt{\frac{\mu 
}{D}}\tan ^{-1}\left( \sqrt{\frac{\mu }{D}}\psi \right) \right] \,.
\label{W}
\end{equation}%
The extrema of the effective potential $W(\psi )$ are the same than those of 
$V(\psi )$, namely $\psi _{k}=2\cos (\theta _{k})/\sqrt{3}$, with $\theta
_{k}=\frac{1}{3}[\cos ^{-1}(-\frac{1}{2}\sqrt{27}h)+2k\pi ]$, and $k=0,1,2$,
see inset to Fig. \ref{vidafi} and Fig. \ref{wpot}. For $h<0$, $\psi _{0}$, $\psi _{1}$ and $%
\psi _{2}$ correspond to the metastable, stable and unstable extrema,
respectively, and the escape time from the metastable minimum is \cite{vanK} 
\begin{equation}
\tau (D,\mu ,h)\approx \frac{2\pi }{\sqrt{|V^{\prime \prime }(\psi
_{0})V^{\prime \prime }(\psi _{2})|}}\,\,\text{exp}\left[ \frac{1}{D}(W(\psi
_{2})-W(\psi _{0}))\right] \,,  \label{vidalange}
\end{equation}%
where $V^{\prime \prime }=\partial _{x}^{2}V(x)$. Identifying $D$ with
the temperature parameter in our microscopic model, this precisely
reproduces the thermal NES phenomenon for $\tau $ as described above;
compare Figs. \ref{vidafi} and \ref{vida}. Therefore, the resonant,
temperature--enhanced stabilization of nonequilibrium metastable states may
be regarded as a consequence of competition between additive and
multiplicative noise.

\subsection{The Limit of Metastability\label{35}}

When the magnetic field $|h|$ is increased, the \emph{strength} of the
metastable state decreases. That is, the local minimum in the free--energy
functional associated to the metastable phase becomes less pronounced. Upon
further increasing $|h|$, such local minimum eventually disappears. At this
point, the metastable state becomes unstable, which means that its
relaxation toward the final stable configuration is no longer hampered by
any free--energy barrier. The magnetic field strength $h^{\ast }(T,p)$ at
which this metastable--unstable transition occurs is known as the spinodal \cite{Klein}
or, sometimes (see Appendix A) pseudospinodal or \emph{intrinsic--coercive
field}.

\begin{figure}[t]
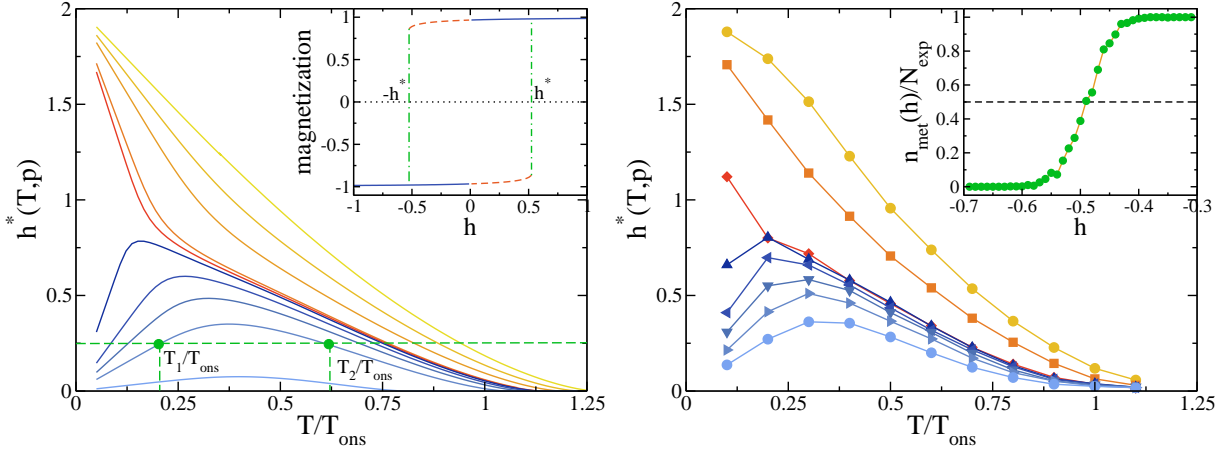

\centerline{
\psfig{file=h-critico-TEORIA-JSP.eps,width=8cm}
\psfig{file=h-critico-MC-JSP2.eps,width=8cm}
}
\caption{{\protect\small (Color) Left panel: Mean--field result for $h^{\ast
}(T,p)$ as a function of $T$ for, from top to bottom, $p=0$, $0.01$, $0.02$, 
$0.03$, $0.031$, $0.032$, $0.035$, $0.04$, $0.05$ and $0.1$. The qualitative
change of behavior in the low temperature region occurs for $p\in
(0.031,0.032)$. We also indicate for $|h|=0.25$ the temperatures $T_{1}<T_{2}
$ which, as discussed in the main text, comprise the metastable region for $%
p=0.05$. The inset shows the two locally--stable steady magnetization
branches as a function of $h$ for $T=0.7T_{c}(0)$ and $p=0.005$. The solid
(dashed) line represents stable (metastable) states. The dot--dashed line
signals the discontinuous transition, at $h=h^{\ast }(T,p)$, where
metastable states disappear. Right panel: Monte Carlo results for $h^{\ast
}(T,p)$ as a function of $T$ for $L=53$ and, from top to bottom, $p=0$, $0.01
$, $0.03$, $0.0305$, $0.0320$, $0.0350$, $0.04$ and $0.05$. The change of
asymptotic behavior in the low-$T$ limit happens for $p\in (0.03,0.0305)$.
The inset shows the probability of occurrence of metastability, as defined
in the main text, as a function of $h<0$ for $L=53$, $T=0.7T_{ons}$ and $p=0.
$ Data here correspond to an average over $500$ independent demagnetization
runs for each value of $h$. }}
\label{spinodal}
\end{figure}

As $|h|$ is increased, the metastable phase remains homogeneous on a
coarse--grained length scale, for times shorter than $\tau ,$ so that the
mean--field approach in section \ref{31} is suitable within this context.
The locally--stable steady magnetization, as computed from the stationary
solution of eqs. (\ref{xzpair1}) and (\ref{xzpair2}), exhibits two branches
as a function of $h$, see the inset in the left graph of Fig. \ref{spinodal}%
. This hysteresis loop reveals that, at the mean--field level, metastability
disappears \emph{abruptly} for any $\left\vert h\right\vert >h^{\ast }$. In
order to evaluate $h^{\ast }(T,p),$ one may study how the metastable state
responds to small perturbations of the applied field. If $(x_{\text{st}}^{%
\text{h}_{0}},z_{\text{st}}^{\text{h}_{0}})$ is a locally--stable stationary
state for $T$, $p$ and $h_{0}$, with magnetization opposed to $h_{0}$, and
we perturb $h=h_{0}+\delta h$, the new locally--stable stationary solution
is modified according to $x_{\text{st}}^{\text{h}}=x_{\text{st}}^{\text{h}%
_{0}}+\epsilon _{\text{x}}$ and $z_{\text{st}}^{\text{h}}=z_{\text{st}}^{%
\text{h}_{0}}+\epsilon _{\text{z}}$. One obtains to first order that 
\begin{equation}
\epsilon _{\text{x}}=\left[ \frac{\partial _{h}G_{2}\partial
_{z}G_{1}-\partial _{h}G_{1}\partial _{z}G_{2}}{\partial _{x}G_{1}\partial
_{z}G_{2}-\partial _{x}G_{2}\partial _{z}G_{1}}\right] _{0}\times \delta h\,,
\label{response}
\end{equation}%
and a similar expression for $\epsilon _{\text{z}}$, where the quantity in
brackets is evaluated at $(x_{\text{st}}^{\text{h}_{0}},z_{\text{st}}^{\text{%
h}_{0}})$ for given values of the parameters $T$, $p$, and $h_{0}$, and the
functions $G_{1,2}(x,z)$ are defined in eqs. (\ref{xzpair1}) and (\ref%
{xzpair2}). This (linear) response diverges for 
\begin{equation}
\left[ \partial _{x}G_{1}\partial _{z}G_{2}-\partial _{x}G_{2}\partial
_{z}G_{1}\right] _{0}=0\,,  \label{condh}
\end{equation}%
which corresponds to a discontinuity in the metastable magnetization as a
function of $h$. For fixed $T$ and $p$, the field for which eq. (\ref{condh}%
) holds is $h^{\ast }$. The left graph of Fig. \ref{spinodal} shows the
mean--field result for $h^{\ast }(T,p)$ as obtained numerically from eq. (%
\ref{condh}). In particular, for $p=0,$ we recover the standard equilibrium
mean--field spinodal curve: Converging to $2$ as $T\rightarrow 0$, linearly
decreasing with temperature for small $T$, and vanishing as $(T_{\text{bethe}%
}-T)^{3/2}$ at the mean--field equilibrium critical point. For $p>0,$ an
instability separates two different low--$T$ behaviors of $h^{\ast }(T,p)$
depending on the amplitude of the nonequilibrium fluctuations. For small
enough values of $p,$ namely, $p\in \lbrack 0,0.031],$ which includes the
equilibrium case, the field $h^{\ast }(T,p)$ monotonously grows and
extrapolates to $2$ as $T\rightarrow 0$. For larger $p$, up to $p=5/32,$
however, $h^{\ast }(T,p)\rightarrow 0$ as $T\rightarrow 0$, with a maximum
at some intermediate value of $T$. 

The value $p=\pi _{c}\approx 0.0315$ separates these two types of asymptotic behavior.
The $p<\pi _{c}$ regime can be understood on simple grounds. In this case, $%
h^{\ast }(T,p)$ increases as $T$ drops, meaning that a stronger field is
needed to \emph{destroy} metastability. In a metastable state, the tendency
to maintain spin order prevails over the tendency of the individual spins
to follow the external field. Since both $T$ and $p$ induce disorder, one
should expect that decreasing $T$ and/or $p$ a stronger field will be needed
to destroy the metastable state, as confirmed in Fig. \ref{spinodal} for $%
p<\pi _{c}$.

The $p>\pi _{c}$ regime is more intriguing. Consider, for instance, $%
p=0.05>\pi _{c}$ and $|h|=0.25$. According to the left graph in Fig. \ref%
{spinodal}, we can define two temperatures such that metastability only
occurs for $T\in \lbrack T_{1},T_{2}]$. The fact that $h^{\ast }$ goes to
zero as $T\rightarrow 0$ in this regime means that, at low temperature, the
nonequilibrium fluctuations are strong enough to \emph{destroy} on their own
the metastable phase. Based on the above argument, one would expect that
increasing $T$ should avoid metastability. However, there is metastability
for intermediate temperatures, $T\in \lbrack T_{1},T_{2}]$. This reentrant
behavior of the spinodal field suggests once more a resonance between
thermal and nonequilibrium fluctuations: despite the intrinsic disordering effect of both noise sources, $T$ and $p$,
they \emph{cooperate} to produce metastability, for which local order prevails. We measured $h^{\ast }$ in MC
simulations, confirming this; see the right graph in Fig. \ref{spinodal} and
appendix A. In particular, we find the instability for $p=\pi _{c}\approx
0.03025$ in simulations, which is very close to the mean--field result.

There is an important difference between the two resonant phenomena above,
i.e., the noise--enhanced stabilization of nonequilibrium metastable states and 
the reentrant behavior of the spinodal field. That is, while the former occurs for 
any $p>0$, the latter only occurs for $p>\pi _{c}$. In order to
understand this fact, we studied the limit of metastability within the
cluster nucleation scheme of section \ref{33}. In this case, $h^{\ast }(T,p)$
may be estimated from the condition that the critical cluster contains one
single spin, $2\mathcal{R}_{c}(T,p,h^{\ast })=1$. This is equivalent to
require that any microscopic fluctuation can trigger the transition toward
the stable state, which signals the onset of instability. The result,
namely, $h^{\ast }(T,p)=\sigma _{0}/m_{0}$ according to eq. (\ref{gotacrit}%
), must be taken with some care, since it assumes that the surface tension 
$\sigma _{0}$, a macroscopic concept, is relevant for microscopic clusters.
In fact, this corresponds to a mean--field spinodal \cite{Rikvold}, which exhibits
reentrant behavior \emph{for any} $p>0$. 
Notice that the nonequilibrium perturbation in our model is stochastic in origin, producing
random spin flips with probability $p$. For a macroscopically large interface the probability that these stochastic 
perturbations do not affect its structure and dynamics is negligible, so the nonequilibrium
perturbations are always present, inducing the observed low--$T$ anomaly in 
$\sigma _{0}$, see $\S $\ref{32}. On the other hand, microscopic clusters (as those
involved in the computation of $h^{\ast }$) have a non--negligible
probability of not being affected by the stochastic perturbation $p$ during their
nucleation and evolution for small enough $p$. In this way, these small
clusters unaffected by $p$ would behave effectively as \emph{equilibrium}
clusters. This would renormalize the (microscopic) critical cluster size to a new, larger
effective value, which thus would modify our estimation of the spinodal field 
$h^{\ast }$. Hence one expects $h^{\ast }$ to behave \emph{as in equilibrium}
for small enough values of $p$, as is in fact observed, while for large
values of $p$ (say $p>\pi _{c}$) even small clusters are affected by the
nonequilibrium perturbation, involving reentrant behavior for $h^{\ast }$.
The challenge remains to put this qualitative argument, which seems
plausible, into more quantitative grounds.

\section{Small Particles\label{sect4}}

\begin{figure}[t]
\centerline{
\psfig{file=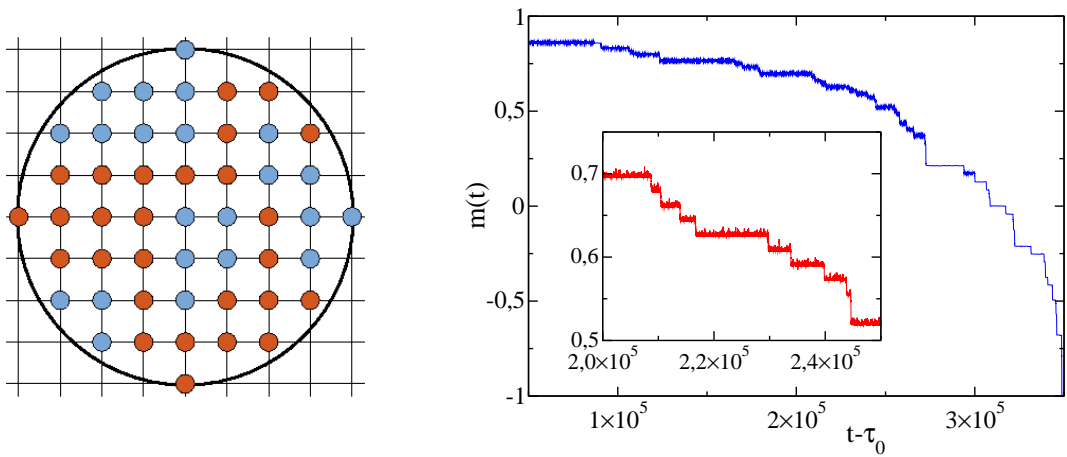,width=6cm}
\psfig{file=JSM-fig1-new2.eps,width=8cm}
}
\caption{{\protect\small (Color) Left: Schematic representation of the model
particle with open circular boundary conditions. Spins at the border do not
have nearest neighbor spins outside the circle. Right: Typical evolution in
which the magnetization is observed to decay by jumps to the final stable
state. This is for a single particle of radius $r=30$ ($\sim 10^{3}$ spins)
at low temperature, and for small values of $h$ and $p$ (see the main text).
The time axis shows $t-\protect\tau _{0}$ in MCSS (Monte Carlo steps per
site) with $\protect\tau _{0}=10^{30}$MCSS; this is of order of the duration
of the initial metastable state. The inset shows a significant detail of the
relaxation. }}
\label{evol}
\end{figure}

The above concerns \emph{bulk} metastability, i.e., a property of
macroscopic systems or models in the thermodynamic limit. However, the case
of finite systems, in which surface effects are relevant, is of great
interest. This occurs for magnetic particles in many applications, e.g.,
magnetic recording media where one often needs to deal with particle sizes ranging
from mesoscopic to atomic levels, namely, clusters of $10^{4}$ to $10^{2}$
spins. We therefore also studied our model with open circular boundary
conditions. The system is now defined on a square lattice, where we inscribe
a circle of radius $r$, and assume that the bonds to any spin outside the
circle are broken; see the left graph in Fig. \ref{evol}.

The effects of free borders on the metastable--stable transition have
already been studied in equilibrium systems \cite{nico,contour,Cirillo}. In
this case, the system evolves to the stable state through the \textit{%
heterogeneous} nucleation of one or several critical clusters, which always
develop at the border, where it is energetically favorable. Apart from this,
the properties of the metastable--stable transition do not change
qualitatively as compared to the case of periodic boundaries. In our
nonequilibrium system we also observe this kind of heterogeneous nucleation.
However, in closer inspection, one notices that the randomness of the decay
process is importantly enhanced, which results in some unexpected
phenomenon. 

The main observation is illustrated in the right graph of Fig. \ref{evol}.
That is, when observed on the appropriate time scale, namely, after each MC
step (per spin), the time relaxation of the magnetization, $m\left( t\right)
,$ from the metastable state occurs by \emph{strictly} monotonic changes.
There is thus a sequence of well--defined abrupt jumps of $m(t),$ which we
will name \textit{avalanches} in the following, that resembles familiar
relaxation processes in many complex systems in nature. We checked that this
is a general feature of the model relaxation at low $T.$ However, a too
rapid relaxation and/or domains too diluted and interfaces too diffuse will
tend to obscure the observation. Consequently, it turned out preferable to
deal with small values of temperature, in order to have compact and
well--defined clusters, and with values of $h$ that do not
excessively accelerate the evolution. In addition, the parameter $p$ can take a wide range of values, 
provided that its effects are comparable to those of temperature. A perfect compromise is 
$T=0.11\,T_{\text{ons}}$, $p=10^{-6}$, and $h=-0.1$, and most data below concern these
parameter values.

\subsection{Avalanche Statistics\label{41}}

To be precise, consider an \textit{avalanche} that begins at time $t_{a}$,
when the system magnetization is $m(t_{a})$, and finishes at $t_{b}$. We
define the avalanche duration as $\Delta _{t}\equiv |t_{a}-t_{b}|$ and its
size as $\Delta _{m}\equiv |m(t_{a})-m(t_{b})|$, and we are interested on
the associated respective distributions, $P\left( \Delta _{t}\right) $ and $%
P\left( \Delta _{m}\right) $. It turns out to be most important for the
reported result to remove from the data some trivial \emph{extrinsic} noise 
\cite{BN2}, namely, small thermal events of typical size (see appendix B) 
\begin{equation}
\bar{\Delta}=\left\{ \ln \frac{\displaystyle \left( 1+p\right) \left[ 1+\rm{e}^{ 2\beta
\left\vert h\right\vert } \right] }{\displaystyle p+\rm{e}^{2\beta \left\vert
h\right\vert } }\right\} ^{-1}.  \label{smallevents}
\end{equation}%
These events correspond to the short--length fluctuations that are evident
by direct inspection in the inset of the right graph in figure \ref{evol}.
These also correspond to avalanches that originate at \emph{flat} interfaces
which, at low enough $T,$ have a significant probability to form, as
discussed below. 
\begin{figure}[t]
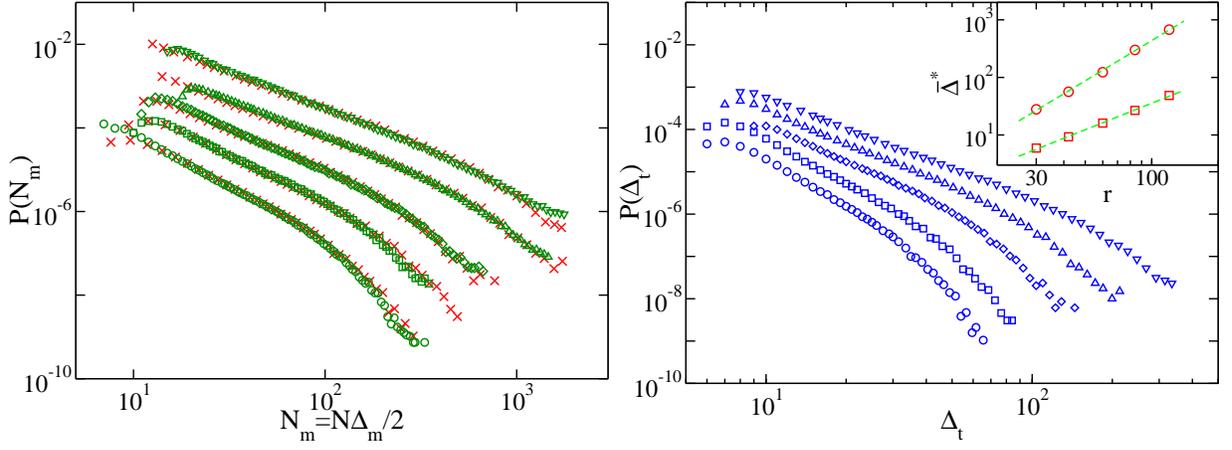

\centerline{
\psfig{file=JSM-fig2-new2.eps,width=8cm}
\psfig{file=JSM-fig3-new3.eps,width=8cm}
}
\caption{{\protect\small (Color) Left: Log--log plot of the size
distribution $P\left( \Delta _{m}\right) $ of \textit{avalanches} (i.e., the
data after subtracting \textquotedblleft small events\textquotedblright\ as
defined in the main text) for ensembles of independent model particles of
radius (from bottom to top) $r=30,$ 42, 60, 84 and $120,$ respectively. This
corresponds to the O symbols (green). Plots of the corresponding duration
distribution $P(\Delta _{t})$ vs. $c\Delta _{t}^{\protect\gamma }$ for each $%
r$ are also shown ($\times ,$ red) superimposed, with $c\approx 0.5$ and $%
\protect\gamma \approx 1.52$ (see the main text). For visual convenience,
the curves are shifted vertically by 4$^{n}$ with $n=0$ to 4 from bottom to
top. Right: Log--log plot of the duration distribution $P\left( \Delta
_{t}\right) $ for the same ensembles. For visual convenience, these curves
are shifted vertically by $2^{n}$ with $n=0$ to 4 from bottom to top. The
inset shows a log--log plot of the size (top) and duration (bottom) cutoffs $%
\bar{\Delta}^{\ast }$ vs. $r$. Lines are power-law fits. Running averages
have been performed in all cases for clarity purposes. }}
\label{size}
\end{figure}

The distribution $P\left( \Delta _{m}\right) $ that results after deducting
these small events is illustrated in the left graph of Fig. \ref{size}. This
is well described by%
\begin{equation}
P\left( \Delta _{m}\right) \sim \Delta _{m}^{-\tau \left( r\right) }\,,
\label{scalm}
\end{equation}%
with a size--dependent exponent given by 
\begin{equation}
\tau \left( r\right) =\tau _{\infty }+a_{1}r^{-2}\,,  \label{scm}
\end{equation}%
where $\tau _{\infty }=1.71\left( 4\right) $. The left graph in Fig. \ref%
{size} depicts the corresponding duration distributions. This follows%
\begin{equation}
P\left( \Delta _{t}\right) \sim \Delta _{t}^{-\alpha \left( r\right) }\,,
\label{scalt}
\end{equation}%
with 
\begin{equation}
\alpha \left( r\right) =\alpha _{\infty }+a_{2}r^{-2}\,,  \label{sct}
\end{equation}%
where $\alpha _{\infty }=2.25\left( 3\right) $. In both cases, size and
duration, the apparent power law ends with an exponential tail, i.e.,%
\begin{equation}
P\left( \Delta \right) \sim \exp \left( -\Delta /\bar{\Delta}^{\ast }\right)
.  \label{cutoff}
\end{equation}%
The cutoffs that we observe follow $\bar{\Delta}^{\ast }\sim r^{\beta }$
with $\beta _{m}\sim 2.32\left( 6\right) $ and $\beta _{t}\sim 1.53\left(
3\right) $, respectively (see the inset in the left graph of Fig. \ref{size}%
).

In practice, we observed this behavior for several combinations of values of
the parameters $T,$ $p$ and $h.$ The requirement for an easy observation is
that, as indicated above, the configurations need to be sufficiently compact
and that none of the underlying processes should predominate so that it
obscures the other. That is, the above behavior is the consequence of a
competition between metastability, thermalization and the nonequilibrium
perturbation. As a matter of fact, as it will be discussed below, observing
power laws requires both free borders and the nonequilibrium condition. That
is, the distributions $P\left( \Delta \right) $ look approximately
exponential if the system has periodic borders and/or one sets $p=0$ in (\ref%
{rate}). It is also obviously convenient, in order to have good statistics,
a choice of the parameter values that selects a region of the phase diagram
in which evolution proceeds by as many jumps as possible. In any case,
however, the behavior that we describe here is robust and characterizes our
model, and no fine--tuning of parameters is needed.

\subsection{Many Scales\label{42}}

We next show that the observed power laws are in the model a superposition
of many individual processes, each with a well--defined scale, and that
these may be traced back to specific demagnetization events. Therefore,
there is no need here to invoke for any critical point, which is the
familiar origin for scale invariance in equilibrium.

\begin{figure}[t]
\centerline{
\psfig{file=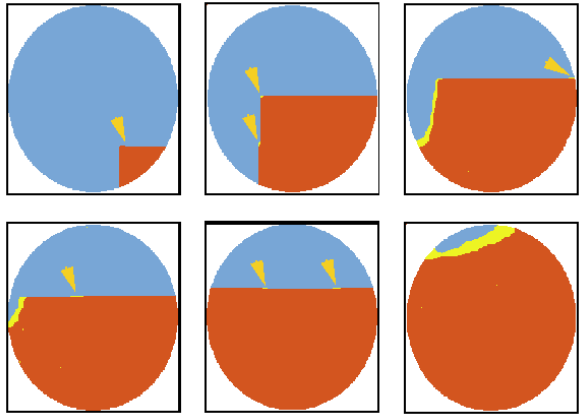,width=8cm}}
\caption{\small (Color) Typical time evolution of a
circular model particle of radius $r=120$ as it decays from the metastable
state. Time increases from left to right and from top to bottom. The
metastable (stable) phase is the blue (red) region, and we represent the
specific avalanches at each time as yellow regions. As discussed in the main
text, all the large avalanches (the only ones that are visible here to the
naked eye; the small others are indicated by arrows) occur at the curved
parts of the interface. This simulation is for $T=0.11\,T_{\text{ons}}$, $%
p=10^{-6}$ and $h=-0.1$.}
\label{snapcirc}
\end{figure}

\begin{figure}[t]
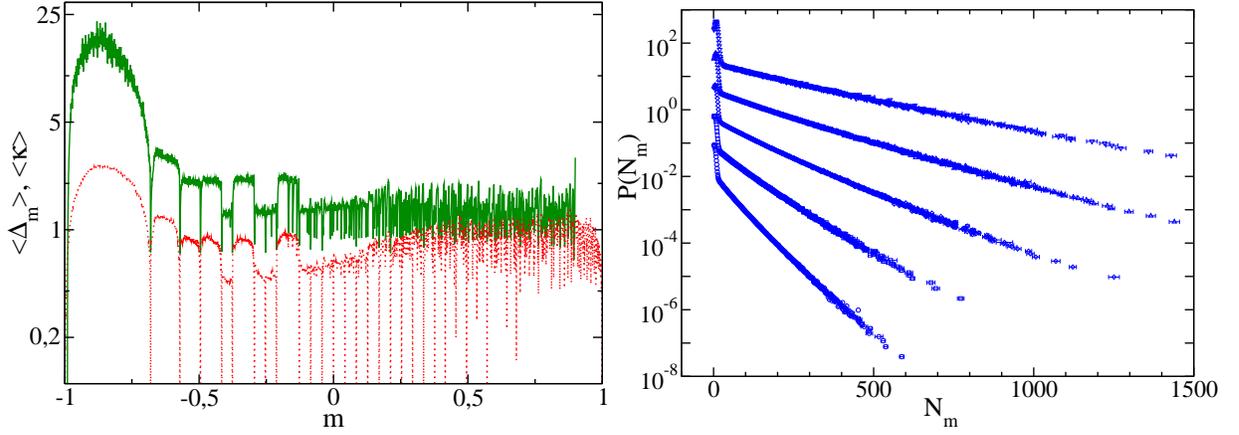

\centerline{
\psfig{file=JSM-fig4-new2.eps,width=8cm}
\psfig{file=JSM-fig5-new2.eps,width=8cm}}
\caption{{\protect\small (Color) Left: Semilogarithmic plot of the mean size 
$\langle \Delta _{m}\rangle $ (solid, green line) and mean curvature $%
\langle \protect\kappa \rangle $ (dotted, red line) as a function of
magnetization. Notice the non--trivial structure uncovering a high degree of
correlation between the avalanche mean size and the average curvature of the
wall at which it originates. Right: Semilogarithmic plot of $P\left( \Delta
_{m}\mid \protect\kappa \right) $, the size distribution for avalanches
developing at a wall of constant curvature, $\protect\kappa ,$ where $%
\protect\kappa $ increases from bottom to top (see appendix C). Here, $%
N_{m}\equiv \frac{1}{2}N\Delta _{m}$. For visual convenience, the curves are
shifted vertically by $10^{n}$ with $n=0$ to 4 from bottom to top. This
corresponds to an average over $3500$ independent runs, and running averages
have also been performed for clarity purposes. }}
\label{curvature}
\end{figure}

Let us consider a relatively--\textit{large} circular model particle.
Inspection of configurations then clearly reveals that, at sufficiently low $T,$ 
the interesting events always occur at the interface between the
metastable and stable phases. One observes curved domain walls
due to the faster growth of the stable phase near the concave open borders.
This is in accordance with a previous conclusion that, in equilibrium, the
critical cluster always nucleates at the free border \cite{contour,Cirillo}.
Any curvature costs energy, so that the large avalanches tend to occur at
curved domain walls, which will tend to produce flatter interfaces; see Fig. 
\ref{snapcirc}. This is confirmed when one monitors the mean avalanche size 
$\left\langle \Delta _{m}\right\rangle $ and the wall curvature $\left\langle
\kappa \right\rangle $ as a function of magnetization $m\left( t\right) .$
We define the curvature $\kappa $ as the number of ascending steps at the
interface. That is, the number of up spins which have two ups and two downs
at the respective sides along the interface. (This definition requires
well--defined compact clusters, which are the rule at low $T.)$ The left
graph in Fig. \ref{curvature} illustrates our results. After averaging over
many runs, definite correlations show up. That is, as one could perhaps have
imagined, the event typical size is determined by the curvature of the
interface region at which the avalanche occurs.

This is also confirmed by monitoring $P\left( \Delta _{m}\mid \kappa \right) 
$, the conditional probability that an avalanche of size $\Delta _{m}$
develops at a wall of curvature $\kappa $. We studied this in detail by
simulating an interface of constant curvature that evolves according to the
transition rate (\ref{rate}); see appendix C. The right graph of Fig. \ref%
{curvature} shows that $P\left( \Delta _{m}\mid \kappa \right) $ has two
regions for given $\kappa $. The first one corresponds to the small events
of typical size (\ref{smallevents}) mentioned above. There is also a region
which, contrary to the situation in Fig. \ref{size}, exhibits (stretched--)
exponential behavior, namely, $P\left( \Delta _{m}\mid \kappa \right) \sim 
\text{exp}\left[ -\left( \Delta _{m}/\bar{\Delta}_{m}\right) ^{\eta }\right] 
$ with $\eta \approx 0.89.$ That is, a region of the interface with
curvature $\kappa $ tends to induce avalanches of typical size $\bar{\Delta}%
_{m}\left( \kappa \right) $.

This fact turns out most relevant because, due to competition between the
randomness induced by free borders and the one induced by $p$ in the
transition rate (\ref{rate}), the interface tends to exhibit a broad range
of curvatures with time, as illustrated in Fig. \ref{snapcirc} and the left
graph in Fig. \ref{curvature}. More specifically, relaxation proceeds via a
series of different configurations, each characterized by a typical
curvature of the interface and by the consequent typical form of the
critical cluster which induces the avalanche. Therefore, what one really
observes when averaging over time is a random combination of many different
avalanches, each with its typical well--defined (gap--separated) size and
duration, which results in an \emph{effective} distribution. The fact that
this combination depicts several decades (more the larger the system is) of
power--law behavior can be understood on simple grounds.

\begin{figure}[t]
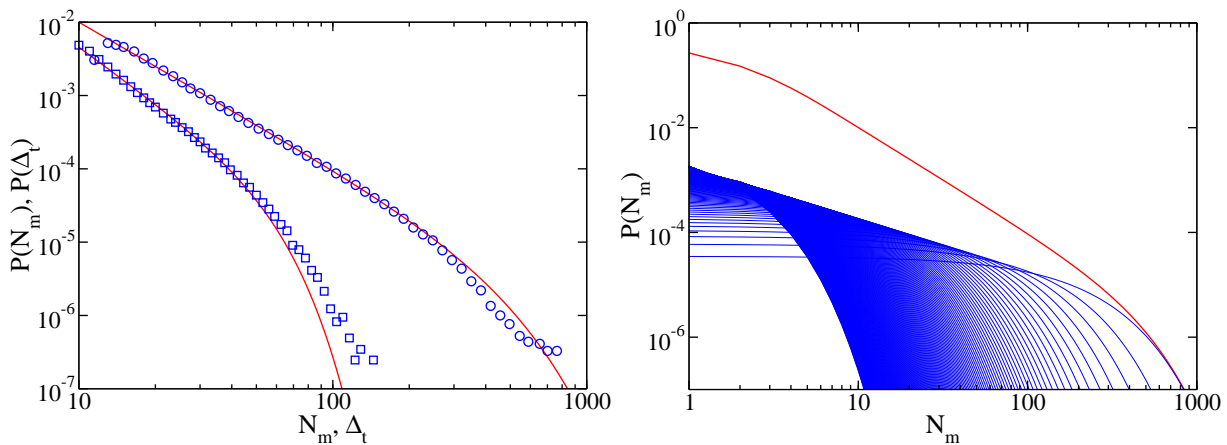

\centerline{
\psfig{file=JSM-fig6-new2.eps,width=8cm}
\psfig{file=exponentials-sum-decomposed.eps,width=8cm}}
\caption{{\protect\small (Color) Left: Solid lines are the predictions, equation (%
\protect\ref{powerdiscreta}) for $n=200,$ $A_{\min }=0.007$ and $A_{\max
}=1. $ The symbols stand for the avalanches duration (lower curve) and size
(upper curve) when $r=60$, i.e., two of the data sets in figures 2 and 3. In
this particular case, the finite--size exponent is $\protect\tau %
(r=60)=2.06(2).$ Right: The theoretical avalanche size distribution on the left panel, and 
the many exponentials which compose it.}}
\label{comparison}
\end{figure}

Let $Q\left( A\right) $ the probability of $A,$ and $P\left( x\mid A\right)
=A\exp \left( -Ax\right) $ the probability of an event of size $x$ given $A$%
. Assume that $A$ can take a finite number of equally spaced values $A_{k}$, 
$k=0,1,2,\ldots ,n$, in the interval $\left[ A_{\min },A_{\max }\right] $,
so that $A_{k}=A_{\min }+k\delta $ with $\delta =\left( A_{\max }-A_{\min
}\right) /n$ (alternatively, one may assume randomly distributed $A_{k}$s),
and that all of them have the same probability, $Q\left( A\right) =\text{%
const}.$ One obtains that 
\begin{equation}
P(x)=\dfrac{\delta \,\text{e}^{-xA_{\min }}}{1-\text{e}^{-x\delta }}\left[
A_{\min }-A_{\max }\,\text{e}^{-(n+1)x\delta }-\delta \,\frac{1-\text{e}%
^{-nx\delta }}{1-\text{e}^{x\delta }}\right] \,.  \label{powerdiscreta}
\end{equation}%
The fact that even such a simple, uncorrelated ansatz describes
qualitatively the data is illustrated in Fig. \ref{comparison}. That is, the
superposition of a large but finite number of exponential distributions,
each with a typical scale, yields an effective global distribution which is
consistent with apparent scale invariance. This distribution extends in
practice up to a cutoff, $\exp \left( -xA_{\min }\right) ,$ which
corresponds to the slowest exponential relaxation. Notice that eq. (\ref%
{powerdiscreta}) predicts a size--independent exponent, $\tau (r)=\tau
_{\infty }=2$, that differs somewhat from the observed asymptotic one, $\tau
_{\infty }=1.71(4)$. This reveals that a more complete explanation than (\ref%
{powerdiscreta}) would be suitable. This would require taking into account,
for example, dynamic correlations such as the ones revealed in the left
graph of Fig. \ref{curvature}. The above, however, is already
semi--quantitative and there is no evidence that a more involved computation
would modify qualitatively this picture.

Consider next $P(\Delta _{t}\mid \Delta _{m})$, i.e., the probability that
the avalanche of size $\Delta _{m}$ lasts a time $\Delta _{t}.$ As expected
from the above, this exhibits well--defined peaks corresponding to large
correlations, i.e., avalanches of a given size have a preferred duration and
vice versa. Assuming $\Delta _{m}\sim \Delta _{t}^{\gamma },$ we obtain $%
\gamma =\beta _{m}/\beta _{t}=1.52(5)$. Using this relation, one may obtain
the duration distribution by combining eq. (\ref{powerdiscreta}) with $%
P(\Delta _{m})\text{d}\Delta _{m}=P(\Delta _{t})\text{d}\Delta _{t}$. A
comparison of the resulting curve with data in Fig. \ref{comparison} leads
to $\gamma \simeq 1.52,$ in perfect agreement with the value obtained from
the cutoff exponents $\beta $. More generally, a scaling plot of $P(\Delta
_{t})$ vs. $c\Delta _{t}^{\gamma }$, with $c$ some proportionality constant,
must collapse onto the corresponding curve $P(\Delta _{m})$ for each
particle radius $r$. This is confirmed in the left graph of Fig. \ref{size}
for $\gamma \simeq 1.52$, further supporting our description of scales
superposition. Let us assume for a moment that, as it would occur in the
presence of a critical condition, both $P(\Delta _{m})$ and $P(\Delta _{t})$
are \textit{true} (as opposed to apparent) power--law distributions. It then
follows the scaling relation $(\alpha -1)=\gamma (\tau -1)$ which, together
with our values above for $\alpha $ and $\tau ,$ implies that $\gamma \simeq
1.76,$ which contradicts the value $\gamma \simeq 1.52$ that follows by two
methods. We believe this misfit simply confirms our point that none of the
distributions $P(\Delta )$ in this section exhibits true scaling behavior.

\subsection{Some Additional Comments\label{43}}

We conclude that the model particle relaxes via events, each with a
well--defined scale, but many of them randomly combine producing (apparent)
power--law behavior. This is a consequence in the model of two main
features, namely, free borders and the nonequilibrium perturbation. (For $p=0
$ and/or periodic boundary conditions, one just observes a well--defined
mean.) There is no indication of chaos, e.g., sensitivity to the initial
condition. We also discard that the observed power laws have any relation in
the model with criticality in the familiar, equilibrium sense. In fact, we
do not observe any relevant correlation other than the dynamic ones that we
described in detail. The question, given that our model is purposely
oversimplified, is whether this picture applies to scale--free fluctuations
in natural phenomena. Demonstrating this, i.e., analysis of separate
elementary events in actual cases, is difficult. However, we argue next that
there are some indications that this may be the case in some occasions.

We first mention that our separation between \textit{small events} ---as
described in eq. (\ref{smallevents}) and appendix B--- and (large) \emph{%
avalanches} is also supported by experiments \cite{BN2,TEB2,Zheng}. On the
other hand, it is remarkable that the statistical properties of the
resulting distributions are indistinguishable in practice from actual data 
\cite{Pablo4}. For instance, size corrections similar to the ones in eqs. (%
\ref{scm}) and (\ref{sct}) for $\tau $ and $\alpha $, respectively, have
been reported in avalanche experiments on rice piles \cite{rice}, and our
asymptotic values are very close to the ones reported in magnetic
experiments, e.g., $\tau _{\infty }=1.77\left( 9\right) ,$ $\alpha _{\infty
}=2.22\left( 8\right) $ and $\gamma =1.51(1)$ in Ref. \cite{BN2} for
(quasi--two dimensional) ribbons; see also Ref. \cite{Zheng}. Moreover, our
cutoff values in eq. (\ref{cutoff}) follow the precise trend observed, for
instance, in magnetic materials \cite{Bahiana,durin}. Our non--critical
scenario is also consistent with the fact that one variously observes either
power--law or exponential distributions, or a mixture of both, in different
but closely related experiments and even in different regimes of the same
experiment; see, for instance, Ref. \cite{jensen}.

\begin{figure}[t!]
\centerline{
\psfig{file=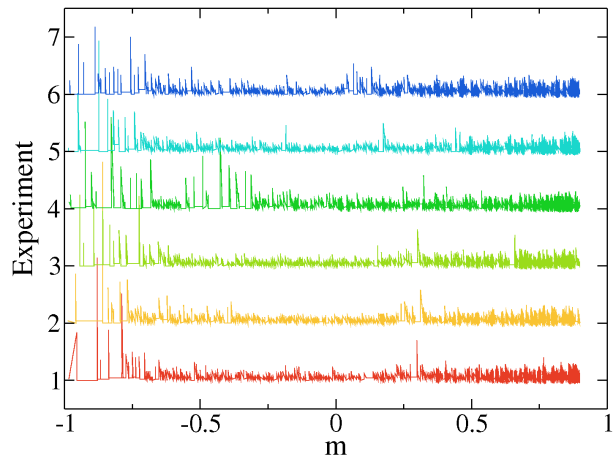,width=8cm}}
\caption
{\small (Color) The avalanche sizes as a function of magnetization for six typical 
independent runs. This is to illustrate that large avalanches tend to occur,
approximately, at the same value of magnetization. This occurs here for the
large avalanche around $m=0.3$ in four of these runs and for the sequence of
large avalanches in all the runs for $m\in \lbrack -0.7,-1],$ for instance.}
\label{repro}
\end{figure}

It is also remarkable that our picture does not imply universality; instead,
microscopic details matter, as it is the case in experiments. That is,
universality is suspicious in a context in which critical exponents seem to
vary with conditions and materials \cite{durin,BN2}, and applications
sometimes rely on sensibility to the sample microstructural details \cite%
{Sipahi0,Sipahi1}. Our picture also explains other features of natural
signals such as, for instance, \textit{reproducibility}. This refers to the
fact that experiments reveal that the \textit{avalanches}, i.e., excluding
the small events, tend to occur at the same stages of evolution \cite%
{weissman,Urbach,extra}. As shown in Fig. \ref{repro}, our system exhibits a
high degree of reproducibility due to the strong tendency of the critical
cluster to nucleate at the border. Simulations indicate that this occurs
more markedly the lower $h$ and $T$ are, i.e., when the system is more
efficient in selecting the most (energetically) favorable configuration.

Interesting enough, if our picture has a broad applicability, there would be more hope to
the goal of predicting large events. That is, the assumption of some
underlying criticality naturally implies that \textit{catastrophes}, though
relatively rare, occur in an strongly correlated bulk and, consequently,
have the same cause as the smaller avalanches \cite{back,econ,sornette}.
Instead, in our picture, events are characterized by their size, and each
size follows from some \textit{specific} microscopic configuration. The
configurations that, under appropriate conditions, may originate large
events qualitatively differ from the ones corresponding to smaller events.
In summary, there is some specific cause for each event which depends on its
size. 

In spite of the extreme simplicity of the model, one should perhaps
recognize that it is likely that its basic features are present in some
actual systems as well. In particular, one may imagine that the model
randomness mimics the causes for the random interface rearrangements in
magnets and for the slip complexity in earthquakes, for instance.
Furthermore, in these two cases, as in the model, time evolution is through
a series of successive short--lived states that constantly halt the system
relaxation. It is precisely this what causes a constant change of scale
which results in the (apparent) scale invariance of the model.

It would be interesting to study the possible occurrence of these
short--lived halt states\ in natural signals. These halts are associated in
the model particle with flat interface regions. That is, once the initial
metastability breaks down, the particle becomes inhomogeneous, and flat
interfaces have a significant probability to form after each avalanche
(which aims at minimizing interfacial energy). As this is the most stable
configuration against small perturbations, the system remains some time with
constant magnetization $m\left( t_{b}\right) .$ This may be described as an 
\textit{entropic metastability.} There is no energy barrier but an unstable
situation such that a \emph{given} microscopic random event suffices to
initiate the next avalanche.

Finally, we mention that there are many other possible explanations for
scale--invariant noise based on non--critical mechanisms; see, for example,
Refs.\cite{weissman,24,bunde,delosR,kaula,davidsen,newman}. To our
knowledge, however, this is the most general one so far reported which has a
physics content. Other proposal are often restricted to specific situations
and, in some cases, they may also be interpreted at the light of a
superposition of many different typical scales. A similar origin for
electronic $1/f$ noises was suggested in the past (see, for instance, Refs. 
\cite{berna,sawa,mazz,proca,weissman}), though this is perhaps the first
time in which an explicit relation is drawn between elementary events
(avalanches) and microscopic physical processes. We also mention that recent
observations in a sociological context may be cast into the behavior of our
\textquotedblleft particle\textquotedblright . That is, according to certain
definitions, there is a relation between fame and merit \cite{fame2}. It has
recently been estimated that the probability distribution for a certain
level of fame falls off exponentially for a respected scientist, while 
it seems to decay as a power law in other social groups \cite{fame2b} This
may reflect that fame in science is mainly due to a homogeneous, rather
well--informed community, while many different communities, each with a
different perception of merit contribute to the fame of, for instance, a
movies actor. The combination of distributions with different means then
produces the observed power law.

\section{Conclusion\label{conc}}

This paper deals with metastability in a nonequilibrium environment. We
studied a two--dimensional Ising ferromagnet subject to competing dynamics,
as if a completely random processes were constantly perturbing a tendency to
thermalization at temperature $T.$ The perturbation impedes equilibrium,
even if it is extremely weak, and a nonequilibrium steady state sets in
asymptotically. This is a convenient background to investigate various
questions, and it could mimic certain nonequilibrium situations in nature.

In the light of the observed qualitative behavior, we argue that, at a
mesoscopic level, concerning both steady and time--dependent properties of
clusters in metastable conditions, it is sensible to assume that a
nonequilibrium \emph{free--energy} function exists which is formally similar
to the equilibrium concept. In particular, we assume that the metastable
phase results from a contest between surface and bulk terms of this
function, and that a nonequilibrium \emph{surface tension} captures the
(strongly nonlinear) interplay between thermal and nonequilibrium \textit{noises}. 
We also show that a Langevin equation with additive and
multiplicative noises captures the essential physics.

Our theoretical approach predicts resonant coupling among the thermal and
nonequilibrium noises which results in novel phenomenology. For instance,
there is noise--enhanced propagation of domain walls and stabilization of
the metastable state at low $T,$ and reentrant behavior of the spinodal
field under strong nonequilibrium conditions. These cooperative phenomena
are perfectly confirmed by Monte Carlo simulations.

We also explored the relaxation from a metastable state in the presence of
open boundaries. This case also shows intriguing behavior at low $T.$ That
is, the decay occurs through a series of many metastable--like states that
repeatedly halt the dynamics, which resembles the relaxation by avalanches
reported for many complex systems, e.g., interface rearrangements in magnets
and fault slips causing earthquakes. We conclude that scale-free avalanche distributions 
are not to be associated to any critical condition in the model but are simply a
superposition of many different classes of events, each with a well--defined
scale. In the model this is determined by the curvature of the domain wall at which
the avalanche originates. This may be a property of any complex relaxation
phenomena characterized by a multiplicity of short--lived, metastable--like
states.

\section*{Acknowledgments}

We acknowledge useful discussions with Miguel \'{A}ngel Mu\~{n}oz, and
financial support from \textit{Ministerio de Educaci\'{o}n y Ciencia, }FEDER
Funds, European Union, and \textit{Junta de Andaluc\'{\i}a}, through
projects FIS2005--00791, HPRN--CT--2002--00307 (DYGLAGEMEM) and FQM--165.

\section*{Appendix A: The Spinodal Field in MC Simulations}

We briefly describe here a method to estimate the spinodal field $h^{\ast}(T,p)$ 
which separates metastable from unstable states. To be precise, $h^{\ast }$ 
is a \emph{pseudo--}spinodal given that the transition in the
actual system (with fluctuations) is not sharp but a progressive crossover \cite{Klein},
as illustrated in the inset to right graph in Fig. \ref{spinodal}.
Estimating $h^{\ast }$ allows in the main text a confirmation of the
corresponding mean--field prediction.

Our method consists in following the system path in phase space as it
evolves from the initial state with all spins up toward the final stable
state. With this aim, we assign to each intermediate state along this path,
say $\mathbf{s}_{k}$, a \emph{measure} $\Lambda (\mathbf{s}_{k})$ of the net
tendency of the system to evolve toward the stable phase from that
configuration. Using the concept of \emph{spin class} described in $\S $\ref%
{31}, it is straightforward to see that $\Lambda (\mathbf{s}_{k})=\mathcal{G}%
(\mathbf{s}_{k})-\mathcal{S}(\mathbf{s}_{k}),$ where%
\begin{equation}
\mathcal{G}(\mathbf{s}_{k})\equiv \sum_{n=0}^{4}\nu (+,n;\mathbf{s}%
_{k})\,\omega (+,n);\qquad \mathcal{S}(\mathbf{s}_{k})\equiv
\sum_{n=0}^{4}\nu (-,n;\mathbf{s}_{k})\,\omega (-,n)  \notag
\end{equation}%
are the growth and shrink rates of the stable phase in state $\mathbf{s}_{k}$%
, respectively (recall that $h<0$). Here, $\nu (s,n;\mathbf{s}_{k})$ is the
fraction of spins in class $(s,n)$ for configuration $\mathbf{s}_{k},$ and $%
\omega (s,n)$ is the spin--class transition rate.

Metastability is hampered by free--energy barriers, while unstable states
evolve without any impediment. Therefore, we may divide relaxation paths in
phase space in two sets: \textit{Metastable paths}, for which at least one
configuration exists such that $\Lambda (\mathbf{s}_{k})<0$, and \textit{%
unstable paths}, where $\Lambda (\mathbf{s}_{k})>0$ $\forall k$. Given the
stochasticity of the dynamics, one needs to be concerned with the \emph{%
probability} of occurrence of metastability, defined as $\Pi (T,p,h)=n_{%
\text{met}}/N_{\text{exp}}$, where $n_{\text{met}}(T,p,h)$ is the number of
experiments out of the total $N_{\text{exp}}$ in which the relaxation path
in phase space belongs to the class of \textit{metastable paths}. The limit
of metastability is defined in this scheme as the field for which $\Pi
(T,p,h^{\ast })=0.5.$ This is illustrated in the right graph of Fig. \ref%
{spinodal} for a system size $L=53.$ (We studied finite--size corrections to
the spinodal field by simulating larger systems; however, these corrections
may be neglected for all practical purposes.) This confirms the reentrant
behavior of $h^{\ast }$ for low temperature under strong nonequilibrium
conditions.

\section*{Appendix B: Small Events}

\begin{figure}[t]
\centerline{
\psfig{file=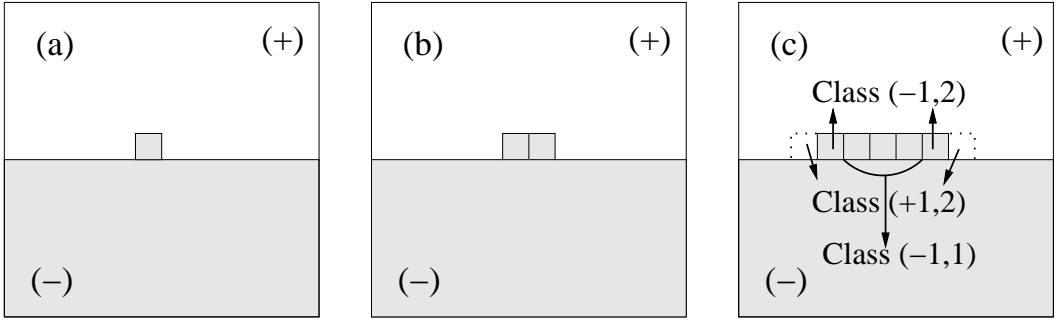, width=14cm}}
\caption{{\protect\small Schematic representation of a flat interface in
which a nucleation event originates and grows as discussed in the main text.
Notice in (c) that the only relevant spins during this kind of growth are
the two spins in class $(-1,2)$ and the two spins in class $(+1,2)$ at the
respective ends of the cluster.}}
\label{avalsmall}
\end{figure}

We now determine the avalanche statistics for a flat interface under a
magnetic field at low $T.$ Let us assume a macroscopic, perfectly \emph{flat}
interface between the metastable and stable phases which, for $h<0,$
correspond to the up and down phases, respectively.

Flat interfaces are very likely at low $T,$ since they minimize surface
tension, and they will tend to invade the metastable phase with time for any 
$h<0.$

We are interested in the probability per unit time for a change in the state
of a spin in class $(s,n),$ $W(s,n),$ where $s=\pm 1$ and $n\in \lbrack 0,4]$
stands for the spin number of up NNs; see $\S $\ref{31}. One has that $%
W(s,n)=\nu (s,n)\,\omega (s,n)$, where $\nu (s,n)$ is the fraction of spins
of class $(s,n)$ in the current configuration and $\omega (s,n)$ is the
corresponding transition rate; see eq. (\ref{rate}) and Table \ref{table}.
For the given initial condition, the only possible classes are $(+1,4),$ $%
(+1,3),$ $(-1,1)$ and $(-1,0).$ For $0\leq \left\vert h\right\vert \leq 2J$
and low $T,$ $\omega (s,n)$ is very small for these four classes, e.g.,
between $10^{-6}$ and $1.24\times 10^{-6}$ for $T=0.11\,T_{\text{ons}}$, $%
p=10^{-6}$ and $h=-0.1.$

Let us assume that a fluctuation nucleates as in Fig. \ref{avalsmall}a,
i.e., one spin has been flipped. This spin belongs to class $(-1,3)$, and
its two NN spins in the direction of the interface belong to class $(+1,2).$
It is straightforward to show that the probability that this fluctuation
either grows along the interface or shrinks is much larger than the
probability of any other event in the surrounding bulk.

Given that there are many more spins in the bulk than in a flat interface,
there will be many bulk fluctuations before anything happens at the
interface. However, whenever an interface fluctuation occurs, an avalanche
will quickly develop along the wall before anything disturbs the surrounding
bulk.

Therefore we can safely assume that once the interfacial perturbation has
appeared, the system dynamics can be reduced to the growth and shrinkage
dynamics of the interfacial perturbation. Under this assumption, the most
probable process to be observed consists in the growth of the interfacial
fluctuation via the flipping of the lateral spins in class $(+1,2)$ at the
extremes, until one of the two spins in class $(-1,2)$ which bound the
interfacial fluctuation flips, halting the avalanche (see Fig. \ref%
{avalsmall}.c). 
\begin{figure}[t]
\centerline{
\psfig{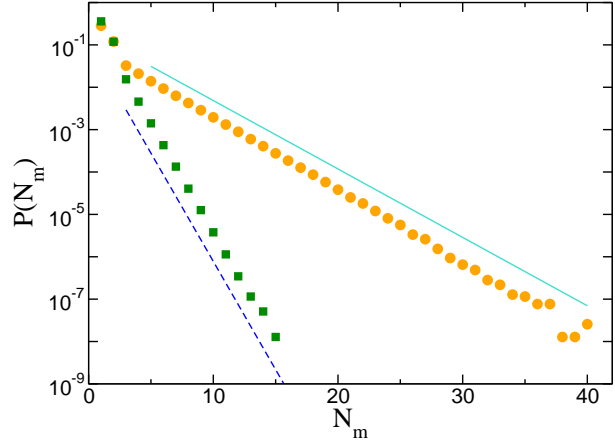}}
\caption{{\protect\small Semilog plot of the avalanche size distributions 
$P(N_{m})$ for avalanches in the field direction ($\bigcirc $) and against
the field ($\square $), as obtained from Monte Carlo simulations of a flat
domain wall of size $L=53$ for $T=0.11\,T_{\text{ons}}$, $p=10^{-6}$ and 
$h=-0.1$. The slope of the lines correspond to our theoretical predictions in
eqs. (\protect\ref{scaledown}) and (\protect\ref{scaleup}). }}
\label{distribupdown}
\end{figure}

We now assume that the system is in the state depicted in Fig. \ref%
{avalsmall}.b, with two up interfacial spins flipped, and we want to compute
the probability of finding a lateral avalanche of size $N_{m}$, i.e.
involving $N_{m}$ spins. The restricted dynamics we assume only involves
four different spins [two in class $(+1,2)$ and two in class $(-1,2)$].
Within our constrained dynamics, the probabilities for the fluctuation to
grow or stop are 
\begin{eqnarray}
\chi _{\text{grow}} &=&\frac{1}{1+p}\left[ p+(1-p)\frac{\text{e}^{2\beta |h|}%
}{1+\text{e}^{2\beta |h|}}\right] \,,  \notag \\
\chi _{\text{stop}} &=&\frac{1}{1+p}\left[ p+(1-p)\frac{\text{e}^{-2\beta
|h|}}{1+\text{e}^{-2\beta |h|}}\right] \,.  \notag
\end{eqnarray}%
The probability of finding a lateral avalanche of size $N_{m}$ is hence $%
P(N_{m})=\chi _{\text{grow}}^{N_{m}}\chi _{\text{stop}}\equiv \chi _{\text{%
stop}}\,\text{exp}(-N_{m}/\bar{\Delta}_{m}^{(-)})$, which defines $\bar{%
\Delta}_{m}^{(-)}$, the typical size characterizing avalanches in the field
direction 
\begin{equation}
\bar{\Delta}_{m}^{(-)}=\frac{1}{\displaystyle\ln \left[ \frac{(1+p)(1+\text{e%
}^{2\beta |h|})}{p+\text{e}^{2\beta |h|}}\right] }\,.  \label{scaledown}
\end{equation}

In a similar way we can compute the typical size $\bar{\Delta}_{m}^{(+)}$ of
avalanches against the magnetic field, i.e. avalanches involving spins in
the stable (down) phase. The result is 
\begin{equation}
\bar{\Delta}_{m}^{(+)}=\frac{1}{\displaystyle\ln \left[ \frac{(1+p)(1+\text{e%
}^{-2\beta |h|})}{p+\text{e}^{-2\beta |h|}}\right] }\,.  \label{scaleup}
\end{equation}%
As expected, $\bar{\Delta}_{m}^{(+)}<\bar{\Delta}_{m}^{(-)}$. Fig. \ref%
{distribupdown} shows the size probability distributions $P(N_{m})$ for
avalanches toward and against the field, as obtained in Monte Carlo
simulations of a flat domain wall of size $L=53$, for for $T=0.11\,T_{\text{%
ons}}$, $p=10^{-6}$ and $h=-0.1$. A comparison with our predictions (\ref%
{scaledown}) and (\ref{scaleup}) is also shown, with excellent results.

The small avalanches here described, to be associated with the presence of
flat domain walls as the system demagnetizes from the initial metastable
state, appear in our magnetic nanoparticle together with more structured
events of large size. As shown in section \ref{sect4}, these structured
events exhibit scale--invariant properties which stem from the interplay
between the nonequilibrium perturbation and the free borders of the magnetic
particle. A good measurement of these large--scale avalanches involves
filtering the above trivial noise, also known as \textit{extrinsic noise} \cite{BN2}.

\section*{Appendix C: Avalanches from Walls of Constant Curvature}

\begin{figure}[t]
\centerline{
\psfig{file=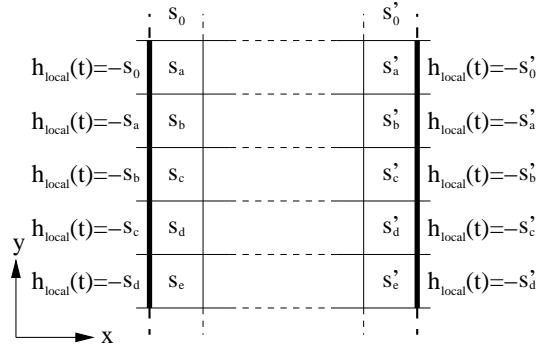,width=7cm}}
\caption{{\protect\small Sketch of the dynamic
boundary conditions in the $\hat{x}$--direction. Here we represent the first
and last columns of the system, and the local, \emph{dynamic} magnetic field
that each spin in these columns feels as the system evolves in time. The
effect of these dynamic boundary conditions is equivalent to eliminate the
interaction of each border spin with its up neighbor, i.e. concave open
boundary conditions. }}
\label{sketch-concavo}
\end{figure}

We have shown in $\S $\ref{42} that large avalanches originate due to domain
wall curvature
\footnote{In this context, we define the amount of curvature of a domain wall as the
number of kinks in the interface, e.g., the number of up spins flanked,
respectively, by two ups and by two downs at the sides along the interface).
This definition requires well--defined compact clusters, as for low
temperature.}.
Moreover, domain wall curvature appear in the system due to 
\emph{concave} open borders, since spins near the concave border flip
faster than bulk spins.

The MC simulations reported in $\S $\ref{42} show that the size of the
avalanche is strongly correlated with the curvature of the interface region
at which the event originates. Here we deep on this correlation, namely, the
interest is on the probability that an avalanche of size $N_{m}$ originates
at a wall of some specific curvature. With this aim, we modified our basic
system so that it shows an interface with constant, except for small
fluctuations, non--zero curvature.

\begin{figure}[t]
\centerline{
\psfig{file=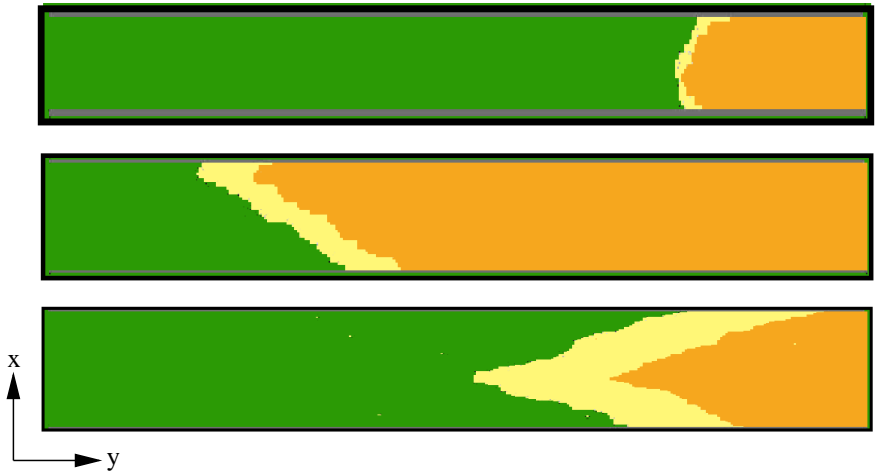,width=12cm}}
\caption
{\small Snapshots of an interface that evolves
according to transition rate (\protect\ref{rate}) subject to \emph{concave}
open boundaries as described in the text. The stable (metastable) phase
corresponds to the green (orange) regions. The yellow regions correspond to
the avalanches in light yellow. From top to bottom, $L_{\text{x}}=20$, $40$,
and $100$.}
\label{snapcurv}
\end{figure}

Consider a semi--infinite, $L_{\text{x}}\times \infty $ lattice with concave
open boundary conditions in the $\hat{x}$ direction. This is done in
practice starting with a $L_{\text{x}}\times L_{\text{y}}$ lattice and
fixing the spins at the top row to $+1$, while spins in the lowest row are
fixed to $-1$. On the other hand, boundary conditions in the $\hat{x}$
direction are \emph{dynamic}: the lattice is also open in the $\hat{x}$
direction, although each spin in the first and last column suffers an
additional \emph{dynamic} magnetic field, equal at any time to the negative
value of its up neighbor. This is sketched in Fig. \ref{sketch-concavo}. For
each spin in the first and the last columns, the effect of these dynamic
boundary conditions is to effectively decouple this spin from its up
neighbor. In this way we mimic a concave, stair--like border (as the one
found by the domain wall in the circular nanoparticle at some stages of its
evolution, see Fig. \ref{snapcirc}), with a fixed distance $L_{\text{x}}$
between both concave borders.

We initialize the system with a flat domain wall between the metastable and
stable phases. Under the action of the magnetic field, the domain wall
propagates toward the metastable phase. Due to the \emph{concave} boundary
conditions in the $\hat{x}$ direction, the interface propagates faster near
the boundary. 
After a short transient, the initially flat domain wall reaches an
stationary state, with an almost constant (up to small fluctuations)
non--zero curvature, which depends linearly on the size $L_{\text{x}}$, see
Fig. \ref{snapcurv}. In order to emulate an infinite system in the $\hat{y}$
direction, we shift the observation window whenever the domain wall gets 
\emph{close} to the system extreme in the propagation direction, always
keeping the interface inside the system. In practice we generate a new
region with up (metastable) spins in one extreme of the system whenever the
above condition is met, eliminating an equivalent region of down (stable)
spins in the opposite extreme. (In fact, in order to give time for the newly
introduced spin region to relax to the typical state of the metastable phase
at the given $T$, $p$ and $h$, we perform the shift well in advance before
the domain wall reaches the boundary.).

The above defined system evolves via avalanches, whose distribution we can
measure. In this way we obtain a reliable measure of avalanche statistics
for a domain wall with (approximately) constant curvature, see the right
graph in Fig. \ref{curvature}. In particular, we observe that domain walls
with constant curvature exhibit avalanches with a well--defined typical
scale.

\end{document}